\documentclass[12pt,a4paper]{article}
\pdfoutput=1

\usepackage{ifthen} 
\usepackage{bigints}
\usepackage{ragged2e}
\newboolean{pdflatex}
\setboolean{pdflatex}{true} 

\newboolean{articletitles}
\setboolean{articletitles}{true} 

\newboolean{uprightparticles}
\setboolean{uprightparticles}{false} 

\newboolean{inbibliography}
\setboolean{inbibliography}{false} 

\usepackage[top=1in, bottom=1.25in, left=1in, right=1in]{geometry}

\usepackage{microtype}
\usepackage{lineno}  
\usepackage{xspace} 
\usepackage{caption} 

\usepackage{graphicx}  
\usepackage{color}
\usepackage{colortbl}
\graphicspath{{./figs/}} 

\usepackage{amsmath} 
\usepackage{amssymb}
\usepackage{amsfonts}
\usepackage{booktabs}
\usepackage{upgreek} 

\newcommand*\patchAmsMathEnvironmentForLineno[1]{%
\expandafter\let\csname old#1\expandafter\endcsname\csname #1\endcsname
\expandafter\let\csname oldend#1\expandafter\endcsname\csname
end#1\endcsname
 \renewenvironment{#1}%
   {\linenomath\csname old#1\endcsname}%
   {\csname oldend#1\endcsname\endlinenomath}%
}
\newcommand*\patchBothAmsMathEnvironmentsForLineno[1]{%
  \patchAmsMathEnvironmentForLineno{#1}%
  \patchAmsMathEnvironmentForLineno{#1*}%
}
\AtBeginDocument{%
\patchBothAmsMathEnvironmentsForLineno{equation}%
\patchBothAmsMathEnvironmentsForLineno{align}%
\patchBothAmsMathEnvironmentsForLineno{flalign}%
\patchBothAmsMathEnvironmentsForLineno{alignat}%
\patchBothAmsMathEnvironmentsForLineno{gather}%
\patchBothAmsMathEnvironmentsForLineno{multline}%
\patchBothAmsMathEnvironmentsForLineno{eqnarray}%
}

\usepackage{hyperref}    
\usepackage[all]{hypcap} 

\definecolor{heartgold}{rgb}{0.5, 0.5, 0.0}

 \def\Pi      {\ensuremath{i}\xspace}

 \def\Pp      {\ensuremath{p}\xspace}

\def\CP                {\ensuremath{C\!P}\xspace}

\def\Bbar    {\kern 0.18em\overline{\kern -0.18em B}{}\xspace}

\mathchardef\PLambda="7103

\newcommand{\Kspizee}{\ensuremath{\KS\to\pi^0e^+e^-}\xspace}
\newcommand{\Kspizmm}{\ensuremath{\KS\to\pi^0\mu^+\mu^-}\xspace}
\newcommand{\Ksgmm}{\ensuremath{\KS\to\gamma\mu^+\mu^-}\xspace}
\newcommand{\Kspipiee}{\ensuremath{\KS\to\pi^+\pi^-e^+e^-}\xspace}
\newcommand{\Ksmmee}{\ensuremath{\KS\to\mu^+\mu^-e^+e^-}\xspace}

\newcommand{\Klmm}{\ensuremath{\KL\to\mu^+\mu^-}\xspace}

\newcommand{\Klpizmm}{\ensuremath{\KL\to\pi^0\mu^+\mu^-}\xspace}

\newcommand{\KpTpi}{\ensuremath{K^+\to\pi^+\pi^+\pi^-}\xspace}
\newcommand{\Kppimumu}{\ensuremath{K^+\to\pi^+\mu^+\mu^-}\xspace}

\newcommand{\Kspipi}{\ensuremath{\KS\to\pi^+\pi^-}\xspace}
\newcommand{\Ksmm}{\ensuremath{\KS\to\mu^+\mu^-}\xspace}
\newcommand{\Ksmmmm}{\ensuremath{\KS\to\mu^+\mu^-\mu^+\mu^-}\xspace}

\newcommand{\Lppi}{\ensuremath{\Lambda\to p\pi^-}\xspace}
\newcommand{\Lppiee}{\ensuremath{\Lambda\to p\pi^-e^+e^-}\xspace}
\newcommand{\Lpmunu}{\ensuremath{\Lambda\to p\mu^-\bar{\nu_{\mu}}}\xspace}

\newcommand{\XiSmunu}{\ensuremath{\Xi^-\to\Sigma^0\mu^-\bar{\nu_{\mu}}}\xspace}
\newcommand{\XiLmunu}{\ensuremath{\Xi^-\to\Lambda\mu^-\bar{\nu_{\mu}}}\xspace}
\newcommand{\XiLpi}{\ensuremath{\Xi^-\to\Lambda\pi^-}\xspace}

\newcommand{\invfb}{\ensuremath{\text{fb}^{-1}}\xspace}

\def\Kbar  {\kern 0.2em\overline{\kern -0.2em K}{}\xspace}

\newcommand{\BRof}[1]{\ensuremath{{\cal B}(#1)}\xspace}

\newcommand{\tev}{\ifthenelse{\boolean{inbibliography}}{\ensuremath{~T\kern -0.05em eV}\xspace}{\ensuremath{\mathrm{\,Te\kern -0.1em V}}}\xspace}
\newcommand{\gev}{\ensuremath{\mathrm{\,Ge\kern -0.1em V}}\xspace}
\newcommand{\mev}{\ensuremath{\mathrm{\,Me\kern -0.1em V}}\xspace}
\newcommand{\kev}{\ensuremath{\mathrm{\,ke\kern -0.1em V}}\xspace}
\newcommand{\ev}{\ensuremath{\mathrm{\,e\kern -0.1em V}}\xspace}
\newcommand{\gevc}{\ensuremath{{\mathrm{\,Ge\kern -0.1em V\!/}c}}\xspace}
\newcommand{\mevc}{\ensuremath{{\mathrm{\,Me\kern -0.1em V\!/}c}}\xspace}
\newcommand{\gevcc}{\ensuremath{{\mathrm{\,Ge\kern -0.1em V\!/}c^2}}\xspace}
\newcommand{\gevgevcccc}{\ensuremath{{\mathrm{\,Ge\kern -0.1em V^2\!/}c^4}}\xspace}
\newcommand{\mevcc}{\ensuremath{{\mathrm{\,Me\kern -0.1em V\!/}c^2}}\xspace}

\newcommand{\Kp}{\ensuremath{K^+}\xspace}
\newcommand{\KS  }{\ensuremath{K^0_{\mathrm{\scriptscriptstyle S}}}\xspace} 
\newcommand{\KL  }{\ensuremath{K^0_{\mathrm{\scriptscriptstyle L}}}\xspace} 
\newcommand{\Kz  }{\ensuremath{K^0\xspace} }

\def\sigmapmumu{\ensuremath{\Sigma^+ \to \Pp \mu^+ \mu^-}\xspace}

\def\sigmapgamma{\ensuremath{\Sigma^+ \to \Pp \gamma}\xspace}

\def\sigmapmumulfv{\ensuremath{\Sigma^+ \to \antiproton \mu^+ \mu^+}\xspace}

\def\sigmappiz{\ensuremath{\Sigma^+ \to \Pp \pi^0}\xspace}
\def\sigmapgamma{\ensuremath{\Sigma^+ \to \Pp \gamma}\xspace}

\def\sigmapee{\ensuremath{\Sigma^+ \to \Pp e^+ e^-}\xspace}
\def\sigmapemu{\ensuremath{\Sigma^+ \to \Pp e^\pm mu^\mp}\xspace}
\def\kplus{\ensuremath{K^+}\xspace}

\def\lambdappiee{\ensuremath{\PLambda^0 \to \Pp \pi^- e^+ e^-}\xspace}
\def\kpiee{\ensuremath{K^+ \to \pi^+ e^+ e^-}\xspace}
\def\Omegalambdapi{\ensuremath{\Omega^- \to \PLambda \pi^-}\xspace}
\def\Xippipi{\ensuremath{\Xi^- \to \Pp \pi^- \pi^-}\xspace}
\def\Xizeroppi{\ensuremath{\Xi^0 \to \Pp \pi^-}\xspace}

\def\ks{\ensuremath{K^0_S}\xspace}

\def\ksmue{\ensuremath{K^0_S \to \mu^+ e^-}\xspace}
\def\klmue{\ensuremath{K^0_L \to \mu^+ e^-}\xspace}
\def\Kpimue{\ensuremath{K^+ \to \pi^+ \mu^+ e^-}\xspace}

\def\kpiee{\ensuremath{K^+ \to \pi^+ e^+ e^- }\xspace}

\def\sigmapmumu{\ensuremath{\Sigma^+ \to p \mu^+ \mu^-}\xspace}

\def\sigmapmumulfv{\ensuremath{\Sigma^+ \to \bar p \mu^+ \mu^+}\xspace}
\def\sigmapee{\ensuremath{\Sigma^+ \to p e^+ e^-}\xspace}
\def\sigmapemu{\ensuremath{\Sigma^+ \to p e^\pm \mu^\mp}\xspace}
\def\sigmappiz{\ensuremath{\Sigma^+ \to p \pi^0}\xspace}

\def\sigmapgamma{\ensuremath{\Sigma^+ \to p \gamma}\xspace}

\def\lambdappiee{\ensuremath{\Lambda \to p \pi^- e^+ e^-}\xspace}

\usepackage{cite} 
\usepackage{mciteplus}

\usepackage{xcolor} 
\definecolor{halfgray}{gray}{0.55} 
\definecolor{webgreen}{rgb}{0,.5,0}
\definecolor{webbrown}{rgb}{.6,0,0}
\definecolor{Maroon}{cmyk}{0, 0.87, 0.68, 0.32}
\definecolor{RoyalBlue}{cmyk}{1, 0.50, 0, 0}

\hypersetup{%
    colorlinks=true, linktocpage=true, pdfstartpage=3, pdfstartview=FitV,%
    breaklinks=true, pdfpagemode=UseNone, pageanchor=true, pdfpagemode=UseOutlines,%
    plainpages=false, bookmarksnumbered, bookmarksopen=true, bookmarksopenlevel=1,%
    urlcolor=webbrown, linkcolor=RoyalBlue, citecolor=webgreen, 
    pdftitle={Prospects for measurements with strange hadrons at LHCb},%
    pdfauthor={},%
    pdfsubject={},%
    pdfkeywords={},%
    pdfcreator={pdfLaTeX},%
  }

\usepackage{wrapfig}

\usepackage{longtable} 
\usepackage[colorinlistoftodos,prependcaption,textsize=tiny]{todonotes}

\begin{document}

\renewcommand{\thefootnote}{\fnsymbol{footnote}}
\setcounter{footnote}{1}

\begin{titlepage}

\vspace*{-1.5cm}

\noindent

\vspace*{4.0cm}

{\normalfont\bfseries\boldmath\huge
\begin{center}
  Prospects for measurements\\ with strange hadrons at LHCb
\end{center}
}

\vspace*{2.0cm}

\begin{center}
  A.~A.~Alves~Junior$^1$,
  M.~O.~Bettler$^2$, 
  A.~Brea~Rodr\'iguez$^1$,
  A.~Casais~Vidal$^1$,
  V.~Chobanova$^1$, 
  X.~Cid~Vidal$^1$, 
  A.~Contu$^3$,
  G.~D'Ambrosio$^4$,
  J.~Dalseno$^1$, 
  F.~Dettori$^5$,  
  V.V.~Gligorov$^6$,
  G.~Graziani$^7$,
  D.~Guadagnoli$^8$,
  T.~Kitahara$^{9,10}$,
  C.~Lazzeroni$^{11}$, 
  M.~Lucio~Mart\'inez$^1$,  
  M.~Moulson$^{12}$,
  C.~Mar\'in~Benito$^{13}$,
  J.~Mart\'in~Camalich$^{14,15}$,
  D.~Mart\'inez~Santos$^1$,
  J.~Prisciandaro $^1$, 
  A.~Puig Navarro$^{16}$, 
  M.~Ramos Pernas$^1$, 
  V.~Renaudin$^{13}$,
  A.~Sergi$^{11}$,
  K.~A.~Zarebski$^{11}$
  \bigskip\\
{\normalfont\itshape\footnotesize
$^1$Instituto Galego de F\'isica de Altas Enerx\'ias (IGFAE), Santiago de Compostela, Spain\\
$^2$Cavendish Laboratory, University of Cambridge, Cambridge, United Kingdom\\
  $^3$INFN Sezione di Cagliari, Cagliari, Italy\\
  $^4$INFN Sezione di Napoli, Napoli, Italy\\
  $^5$Oliver Lodge Laboratory, University of Liverpool, Liverpool, United Kingdom, 
  now at Universit\`{a} degli Studi di Cagliari, Cagliari, Italy\\
  $^6$LPNHE, Sorbonne Universit\'{e}, Universit\'{e} Paris Diderot, CNRS/IN2P3, Paris, France\\
  $^7$INFN Sezione di Firenze, Firenze, Italy\\
  $^8$Laboratoire d'Annecy-le-Vieux de Physique Th\'eorique\,, Annecy Cedex, France\\
  $^9$Institute for Theoretical Particle Physics (TTP), Karlsruhe Institute of Technology, Kalsruhe, Germany\\
  $^{10}$Institute for Nuclear Physics (IKP), Karlsruhe Institute of Technology, Kalsruhe, Germany\\
  $^{11}$School of Physics and Astronomy, University of Birmingham, Birmingham, United Kingdom\\
  $^{12}$INFN Laboratori Nazionali di Frascati, Frascati, Italy\\
  $^{13}$Laboratoire de l'Accelerateur Lineaire (LAL),  Orsay, France \\
  $^{14}$Instituto de Astrof\'isica de Canarias and  Universidad de La Laguna, Departamento de Astrof\'isica, La Laguna, Tenerife, Spain\\
  $^{15}$CERN, CH-1211, Geneva 23, Switzerland \\
  $^{16}$Physik-Institut, Universit{\"a}t Z{\"u}rich, Z{\"u}rich, Switzerland\\

}
\end{center}

\vspace{\fill}

\begin{abstract}
  \noindent
  This report details the capabilities of LHCb and its upgrades towards the study of kaons and hyperons. 
  The analyses performed so far are reviewed, elaborating on the prospects for some key decay channels, 
  while proposing some new measurements in LHCb to expand its strangeness research program.
  
\end{abstract}

\vspace*{2.0cm}
\vspace{\fill}

\end{titlepage}

\pagestyle{empty}  


\newpage
\setcounter{page}{2}
\mbox{~}

\cleardoublepage

\newcommand{\tmpdg}[1]{{\textcolor{red}{[DG: #1]}}}
\renewcommand{\thefootnote}{\arabic{footnote}}
\setcounter{footnote}{0}

\pagestyle{plain} 
\setcounter{page}{1}
\pagenumbering{arabic}

\section{Introduction}
\label{sec:Introduction}
The study of strange-hadron decays has fuelled discoveries in particle physics for the past
seventy years. For instance, experimental anomalies in the strange sector motivated the prediction of the charm quark via the Glashow-Iliopoulos-Maiani (GIM) mechanism, while the discovery of \CP violation prompted the postulation of the beauty and top quarks within the Cabibbo-Kobayashi-Maskawa (CKM) paradigm; all now key ingredients of the Standard Model (SM). Presently, strangeness decays are valuable probes in the search for dynamics
Beyond the Standard Model (BSM), being particularly relevant in searches
for sources of quark flavour violation beyond the CKM matrix. Since $s\to d$
transitions have the strongest suppression factor, they can typically probe energy
scales higher than those accessible in charm or beauty-hadron decays for couplings of comparable
size~\cite{Nir:2007xn}. Nevertheless, flavour physics experiments have greatly enhanced such knowledge
from charm and beauty decays in recent years, while few measurements of strange-hadron
decays have been updated or performed for the first time.

Several dedicated experiments exist for specific measurements, however few
experiments possess the potential to construct a comprehensive program on the study of
strange hadrons. In this work, it is argued that the LHCb experiment has the capacity,
both in terms of detector performance and statistics, to produce leading measurements
exploiting almost all strange-hadron species, particularly in the search for their rare decays.
An
overview of the current results and prospects of strangeness decays at LHCb is given,
demonstrating LHCb's unique reach as a strangeness factory and motivating further research
in this area.
In fact, the LHCb collaboration has already published the world's most precise measurements in
\Ksmm~\cite{LHCb-PAPER-2012-023,LHCb-PAPER-2017-009} and \sigmapmumu
~\cite{LHCb-PAPER-2017-049}, while projecting world-leading results for \Kspizmm~\cite{Chobanova:2195218}
and \Kspipiee ~\cite{MarinBenito:2193358}.
Experiments such as BESIII~\cite{Li:2016tlt}, NA62~\cite{NA62:2312430,NA62:2017rwk}, KLOE2~\cite{AmelinoCamelia:2010me}, 
KOTO~\cite{KOTO,Ahn:2018mvc} and CLAS~\cite{CLAS, CLAS12,Mecking:2003zu} further enrich the field with diverse and complementary research programs of their own.

This document is organised as follows: Section~\ref{sec:rec} is dedicated to the discussion of the production of strange-hadron decays at LHC
and its detection in LHCb. Section~\ref{sec:raredecays} summarises the results and prospects of LHCb for several rare decays of strange hadrons.
The capabilities for the measurement of the $\Kp$ mass as well as for the study of semileptonic hyperon decays are presented in section~\ref{sec:others}, while conclusions are drawn in section~\ref{sec:conclusions}.

\section{Production and detection of strange hadrons}
\label{sec:rec}

The LHCb detector~\cite{Alves:2008zz} is a single-arm forward spectrometer,
covering the pseudorapidity range $2 < \eta < 5$, collecting data in proton-proton collisions
at the Large Hadron Collider at CERN.  
It is composed of a silicon-strip vertex detector surrounding the $pp$ interaction region (VELO),
with a length of about 1 metre from the interaction point, 
a large-area silicon-strip detector (TT) located upstream of a dipole magnet 
and three tracking stations of silicon-strip detectors and straw drift tubes
placed downstream of the magnet. Particle identification is provided by two ring-imaging
Cherenkov detectors, an electromagnetic and a hadronic calorimeter, and a muon system
composed of alternating layers of iron and multiwire proportional chambers.
LHCb has collected so far an integrated luminosity of about 8 \invfb.

The LHCb detector will be upgraded for the next run of the LHC. This upgrade, hereafter referred to as Phase-I, includes a completely
  new tracking system with a pixel-based VELO~\cite{Collaboration:1647400}, the Upstream Tracker (UT) replacing the TT and scintillating fibre
  detectors acting as tracking stations~\cite{Collaboration:1647400}. The Phase-I detector will collect on the order of $50$~\invfb of integrated luminosity~\cite{Bediaga:2012py}.
  An Expression of Interest for a second upgrade, hereafter denoted  as Phase-II, can be found in ref.~\cite{Aaij:2244311}. It is intended that on the order of $300$~\invfb of integrated luminosity will be collected with this upgrade.

\begin{wrapfigure}{R}{0.36\textwidth}
\includegraphics[trim=0cm 2cm 1cm 3.0cm, clip, width = 0.36\textwidth ]{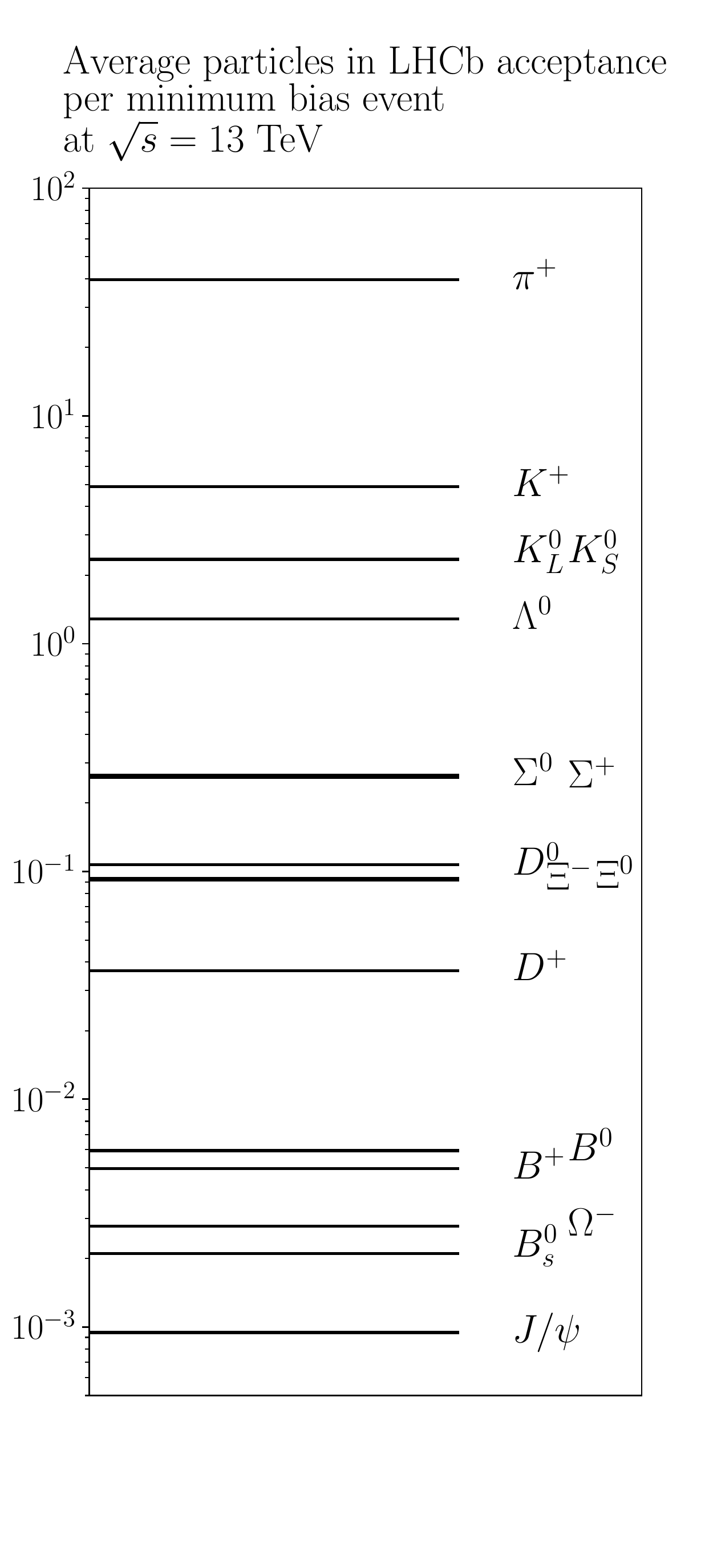}
\caption{Multiplicity of particles produced in a single $pp$ interaction at $\sqrt{s} = 13$~TeV within LHCb acceptance.}\label{fig:minbias}
\vspace{-1cm}
\end{wrapfigure}

The production of strange hadrons at LHC is exceedingly abundant. Physics projections are derived from simulated events invoking the \verb|Pythia| software generator~\cite{Sjostrand:2007gs}, where proton-proton collisions are configured with a centre-of-mass energy $\sqrt{s} = 13$~TeV and 
an average of one interaction per collision. The conclusions of this study are unaffected for other anticipated LHC collision energies of $14$~TeV even up to 28~TeV. 
Multiplicities of various particles are estimated from these events in a broad LHCb 
geometric acceptance of pseudorapidity $\eta \in [1,6]$, prior to any simulated detector response.
This multiplicity is shown for strange hadrons in figure~\ref{fig:minbias} alongside an assortment of well-known heavy flavoured hadrons for comparison. 
Multiple kaons and about one hyperon per event are expected to be produced in these interactions, which is roughly two and three orders of magnitude greater than for charmed and beauty hadrons, respectively. 
Thus, the LHCb experiment will have at its disposal the statistics necessary both for precision measurements of 
strange-hadron decays and for searches for their rare decays.

\begin{figure}[!t]
\begin{center}
    \includegraphics[width= 0.7\textwidth]{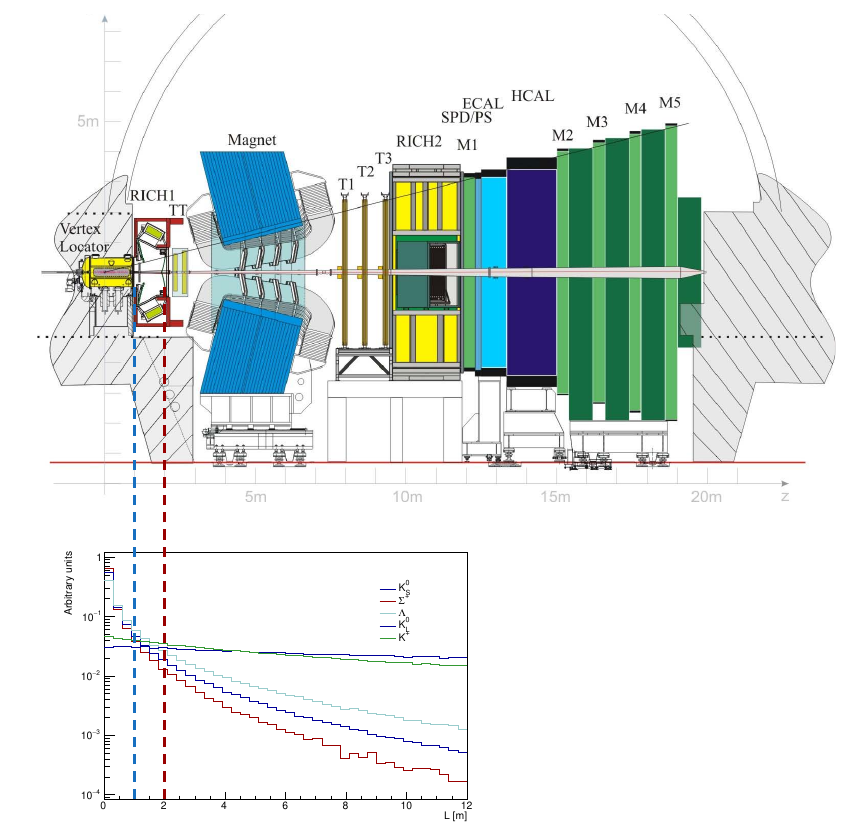}
%
%
    \caption{A side view of the LHCb detector layout~\cite{Alves:2008zz} compared with the decay length of strange hadrons in $pp$ collisions at 
    $\sqrt{s}=13$~TeV. The blue (1m) and red (2m) dashed lines indicate the approximate distance from the interaction point at which daughter tracks can be reconstructed as long and downstream tracks, respectively.}
    \label{fig:lifetime}
\end{center}
\end{figure}

The efficiency of detecting strange-hadron decays will, however, not be the same as for heavy flavour for several reasons.
The detector layout, which is optimised for $b$ decays, implies a relatively lower acceptance for \KS, with \KL and $K^+$ efficiencies diminished even further. This is due to the differing flight lengths of the different mesons. The typical
decay length of a $B$ meson is $\sim 1$ cm, \KS can fly a distance of nearly one metre, while $K^{\pm}$ and \KL traverse distances longer than the full LHCb detector length on average.
Flight distance distributions achieved by various strange hadrons before decaying are also obtained from Pythia simulations,
which are displayed within the context of the LHCb detector in figure~\ref{fig:lifetime}. 

Depending on the decay position of a given particle, its charged decay products
can be reconstructed in LHCb exploiting the relevant tracking sub-detectors. The different track categories are defined in ref.~\cite{Collaboration:1647400} as:
\begin{itemize}
\item long tracks: when all possible tracking information from the VELO to the T stations is available, implying that the mother particle decayed within about 1 metre of the $pp$ interaction point;
  \item downstream tracks: where only the TT and T stations register tracks, allowing strange hadrons to be reconstructed 
  with decay lengths up to about 2 metres from the interaction point.
\end{itemize}
In order to provide an estimate of the reconstruction efficiencies for long tracks, 
the published \Ksmm analysis from LHCb is taken as a benchmark~\cite{LHCb-PAPER-2017-009}.
Events with a decay time $t$ in the range of $t/\tau_{S} \in [0.10, 1.45]$ were used, where $\tau_{S}$ is the \KS lifetime. 
From these numbers, one could simply obtain
\begin{equation}
  \frac{\epsilon_{\KL}}{\epsilon_{\KS}} \approx 3.5\times10^{-3} \quad,\nonumber
\end{equation}
for the ratio of \KL to \KS efficiencies, ${\epsilon_{\KL}}$ and ${\epsilon_{\KS}}$, respectively.

However, as the acceptance inside the VELO is not uniform, larger lifetimes result in lower reconstruction efficiencies, further reducing ${\epsilon_{\KL}}$ next to ${\epsilon_{\KS}}$.
This can be approximated by an exponential acceptance or so-called `beta factor' $\epsilon(t)\sim e^{-\beta t}$~\cite{LHCb-PAPER-2013-002}, with
$\beta \sim 86\,$ns$^{-1}$ in the case of \Ksmm decays~\cite{CidVidal:1490381}.
In this case, the reduction factor becomes
\begin{equation}
  \label{eq:acc2}
  \frac{\epsilon_{\KL}}{\epsilon_{\KS}} = \frac{\Gamma_L\bigint_{0.1\tau_S}^{1.45\tau_S}e^{-t(\Gamma_S+\beta)}dt}{\Gamma_S \bigint_{0.1\tau_S}^{1.45\tau_S}e^{-t(\Gamma_L+\beta)}dt} \approx 2.2 \times 10^{-3},
\end{equation}
\noindent where $\Gamma_S$ and $\Gamma_L$ are the \KS and \KL decay widths.
Assuming that the same acceptance parametrisation used in eq.~\eqref{eq:acc2} holds also for $K^{\pm}$, 
the relative efficiency of $K^{\pm}$ decays with respect to \KS decays is then at the level of 1\%.
On the other hand, the use of downstream tracks can allow for an increased lifetime acceptance. 
The transverse momenta of the products of strangeness decays, significantly softer than for $b$-hadron decays, are also detrimental to their detection at LHCb. 
While $b$-hadron decay products generally have a transverse
momenta of around 1-2~\gevc, for $s$-hadron decays the range is more in the region of 100-200~\mevc. 
The acceptances for several benchmark channels, as well as invariant-mass resolutions, are estimated in the following applying a simplified simulation of the LHCb upgrade tracking, based on the
detector descriptions found in refs.~\cite{Collaboration:1647400,Collaboration:1624070,Hommels:999327}. 
The following selection criteria are applied to all decay channels: the daughter particles are required to have a track impact
parameter to primary vertex of greater than 400~microns, a momentum greater than 3000~\mevc with transverse momentum greater than 80~\mevc, 
while the reconstructed hadron is required to have a decay time greater than 8.9 ps and 
a flight distance in the plane transverse to the beam greater than 3 mm. These requirements are based
on the Run 2 trigger for detached soft dimuons~\cite{Dettori:2297352} and on the lower decay time requirement
from ref.~\cite{LHCb-PAPER-2012-023}. These requirements are expected to be realistic also for future data-acquisition periods in LHCb.
Acceptances are then normalised to that of fully reconstructed \Ksmm, which is found to be around $1\%$.
The results of this simplified simulation are given in table~\ref{tab:acc},
where the efficiency is shown using long tracks only ($\epsilon_L$) and using downstream tracks only ($\epsilon_D$), 
along with the invariant-mass resolution for each reconstruction method.
The efficiency scale factors for charged hadrons with at least 300~\mevc and electrons with over 200~\mevc transverse momenta are also normalised to fully reconstructed \Ksmm and indicated in parentheses. As neutral particles like the photon, neutrino and $\pi^0$ are not reconstructed in this study, the invariant mass of particular strange hadrons is calculated with additional kinematic constraints.

Absolute efficiencies depend significantly on the fidelity of the momentum spectra provided by Pythia,
hence it is preferred to quote only relative acceptances here.
As bremsstrahlung corrections are important in electron reconstruction, 
such modes are shown separately in table~\ref{tab:acc_el}, in which the reference channel for efficiency
normalisation is \Kspipiee. The reconstruction and selection efficiency for \Kspipiee has been estimated
with full LHCb simulation to be $\sim 1\times10^{-4}$ in ref.~\cite{MarinBenito:2193358}. Lepton Flavour Violating
(LFV) modes are listed in table~\ref{tab:acc_lfv}, normalised to \ksmue.

\begin{table}[h]
  \footnotesize
  \centering
  \caption{Acceptance scale factors $\epsilon$, and mass resolutions $\sigma$, for only long~($L$) and only downstream~($D$) tracks obtained from our simplified description of the LHCb Upgrade tracking system geometry. The production ratio of the strange hadron with respect to $K_S^0$ is shown as  $\mathcal{R}$. All efficiencies are normalised to that of fully reconstructed \Ksmm and averaged over particles and anti-particles. Channels containing a photon, neutrino and $\pi^0$ are partially reconstructed.}
  \label{tab:acc}
  \begin{tabular}{lccccc}
  \toprule
    Channel & $\mathcal{R}$ &$\epsilon_{L}$ & $\epsilon_{D}$ & $\sigma_{L}$(\mevcc) & $\sigma_{D}$(\mevcc) \\ \midrule
    \Ksmm & 1 &1.0 (1.0) & 1.8 (1.8) & $\sim 3.0$ & $\sim 8.0$\\ 
    \Kspipi & 1 & 1.1 (0.30) & 1.9 (0.91) & $\sim 2.5$ & $\sim 7.0$\\
    $\Kspizmm$&1 & 0.93 (0.93) & 1.5 (1.5) & $\sim 35$ & $\sim 45$\\
    $\Ksgmm$ &1& 0.85 (0.85) & 1.4 (1.4) & $\sim 60$ & $\sim 60$\\ 
    \Ksmmmm & 1& 0.37 (0.37) & 1.1 (1.1) & $\sim 1.0$ & $\sim 6.0$\\ 
    \Klmm & $\sim 1$ & 2.7 (2.7) $\times 10^{-3}$ & 0.014 (0.014) & $\sim 3.0$ & $\sim 7.0$\\
    \KpTpi & $\sim 2$ & 9.0 (0.75) $\times 10^{-3}$ & 41 (8.6) $\times 10^{-3}$ & $\sim 1.0$ & $\sim 4.0$\\
    \Kppimumu & $\sim 2$ & 6.3 (2.3) $\times 10^{-3}$  & 0.030 (0.014) & $\sim 1.5$ & $\sim 4.5$\\
    \sigmapmumu & $\sim 0.13$ & 0.28 (0.28) & 0.64 (0.64) & $\sim 1.0$ & $\sim 3.0$ \\
    \Lppi & $\sim 0.45$& 0.41 (0.075) & 1.3 (0.39) & $\sim 1.5$ & $\sim 5.0$\\ 
    \Lpmunu &$\sim 0.45$ &  0.32 (0.31) & 0.88 (0.86) & $-$ & $-$\\
    \XiLmunu & $\sim 0.04$ & 39 (5.7) $\times10^{-3}$ & 0.27 (0.09) & $-$ & $-$ \\
    \XiSmunu & $\sim 0.03$ & 24 (4.9) $\times10^{-3}$ & 0.21 (0.068) & $-$ & $-$ \\
    \Xippipi & $\sim 0.03$ & 0.41(0.05) & 0.94 (0.20) & $\sim 3.0$ & $\sim 9.0$ \\
    \Xizeroppi   &$\sim 0.03$&1.0 (0.48) &2.0 (1.3) &$\sim 5.0$&$\sim 10$ \\
    \Omegalambdapi & $\sim 0.001$ & 95 (6.7) $\times10^{-3}$ & 0.32 (0.10) & $\sim7.0$ & $\sim20$ \\
\bottomrule
  \end{tabular}
\end{table}

\begin{table}[h]
  \footnotesize
  \centering
  \caption{Acceptance scale factors $\epsilon$, and mass resolutions $\sigma$, for only long~($L$) and only downstream~($D$) tracks obtained from our simplified description of the LHCb Upgrade tracking system geometry. All efficiencies are normalised to that of fully reconstructed \Kspipiee and are averaged between particles and anti-particles. The invariant-mass resolutions
  shown in the table correspond to the ideal case of perfect bremsstrahlung recovery.}
  \label{tab:acc_el}
  \begin{tabular}{lccccc}
  \toprule
    Channel & $\mathcal{R}$ & $\epsilon_{L}$ & $\epsilon_{D}$ & $\sigma_{L}$(\mevcc) & $\sigma_{D}$(\mevcc) \\ \midrule
    \Kspipiee & 1 & 1.0 (0.18) & 2.83 (1.1) & $\sim 2.0$ & $\sim 10$\\ 
    \Ksmmee &  1 & 1.18 (0.48) & 2.93 (1.4) & $\sim 2.0$ & $\sim 11$\\
    \kpiee       & $\sim 2$  & 0.04 (0.01) & 0.17 (0.06) & $\sim 3.0$ &$\sim 13$\\
    \sigmapee &  $\sim 0.13$  & 1.76 (0.56) & 3.2 (1.3) & $\sim 3.5$ & $\sim 11$\\
    \lambdappiee & $\sim 0.45$   & $<2.2\times10^{-4}$&$\sim 17$ $(<2.2)$ $\times10^{-4}$ & $-$ & $-$ \\
    \bottomrule
  \end{tabular}
\end{table}

\begin{table}[h]
  \footnotesize
  \centering
  \caption{Acceptance scale factors $\epsilon$, and mass resolutions $\sigma$, for only long~($L$) and only downstream~($D$) tracks obtained from our simplified description of the LHCb Upgrade tracking system geometry. All efficiencies are normalised to that of fully reconstructed \ksmue and averaged between particles and anti-particles. The invariant-mass resolutions
  shown in the table correspond to the ideal case of perfect bremsstrahlung recovery.}
  \label{tab:acc_lfv}
  \begin{tabular}{lccccc}
  \toprule
    Channel & $\mathcal{R}$ & $\epsilon_{L}$ & $\epsilon_{D}$ & $\sigma_{L}$(\mevcc) & $\sigma_{D}$(\mevcc) \\ \midrule
    \ksmue & 1 & 1.0 (0.84) & 1.5 (1.3) & $\sim 3.0$ & $\sim 8.0$\\ 
    \klmue &  1 & 3.1 (2.6) $\times 10^{-3}$ & 13 (11) $\times 10^{-3}$ & $\sim 3.0$ & $\sim 7.0$\\
    \Kpimue &  $\sim 2$  & 3.1 (1.1) $\times 10^{-3}$ & 16 (8.5)$\times 10^{-3}$ & $\sim 2.0$ & $\sim 8.0$\\
    \bottomrule
    \end{tabular}
\end{table}

\clearpage
\newpage
\subsection{Trigger}

The current trigger of LHCb has three stages, a hardware stage (L0) and two software stages (HLT1 and HLT2). The L0 is practically unchangeable and
implies an efficiency loss of roughly $80\%$ of reconstructible strange-hadron decays involving muons~\cite{Dettori:2297352}.
For non muonic final states it implies a loss of about 90\% to 99\%, due to the much larger transverse energy trigger thresholds for hadrons and electrons~\cite{Puig:1970930}, 
depending on whether also events triggered by the underlying event (and not by the considered signal) are accepted or not~\cite{Tolk:1701134}.
During Run 1, the total trigger efficiency for strangeness decays was
$\text{1--2}\%$ or lower, depending on the final state. The main reason for those low efficiencies is the soft transverse momentum spectrum of 
strange-hadron decay products.
During Run 2, dedicated software triggers for strange-hadron decays into dimuons have been implemented
with an overall improvement of about one order of magnitude in the total trigger efficiency achieved with respect to Run 1~\cite{Dettori:2297352}. 
In the Upgrade of the LHCb experiment, the trigger is expected to be entirely software
based with L0 removed, hence ${\cal O}(1)$ efficiencies are 
attainable.~\footnote{Here and in the following, trigger efficiencies are calculated and referred to events that have passed the full offline selection, 
hence perfect efficiencies are attainable when the trigger requirements are aligned to, or looser than, the offline selection.\label{footnote:trigger}} 
It has been shown in simulation that for dimuon final states, the output rate can
be kept under control for transverse momentum thresholds as low as $80~\mevc$ without any significant signal loss~\cite{Chobanova:2195218}.
Although the dimuon final state is the cleanest signature from an experimental perspective, trigger algorithms for other final states are possible and are currently
under investigation. As an example, a software trigger for dielectrons from strange decays was already implemented during Run 2~\cite{MarinBenito:2193358} 
and will serve as a basis for the Upgrade. 

\subsection{Flavour Tagging}

As pointed out in ref.~\cite{DAmbrosio:2017klp}, \KS-\KL interference has an effective lifetime which is only twice that of
the \KS and thus has an enhanced acceptance in LHCb compared to pure \KL decays. 
By tagging the initial flavour of the \Kz, access to \KL physics and \CP phenomena in the $\KS-\KL$ system 
is permitted through these interference effects. 
\footnote{While the present paper is focused mainly on rare and semileptonic decays, a program of measurements of \CP violation in the $\KS-\KL$ system
is in principle possible and merits further study.}
Though not used for this paper, it is valuable to mention the possibility of strange-hadron 
flavour tagging at LHCb through \Kz processes such as
$pp\rightarrow K^0 K^- X$, $pp\rightarrow K^{*+}X \rightarrow K^0\pi^+ X$ and $pp\rightarrow K^0\Lambda^0X$.

\section{Rare decays}
\label{sec:raredecays}

Rare decays are excellent probes for BSM. On the theoretical
side, the SM background to each process is small by definition, while experimentally, measurements are typically statistically limited, but this limitation can constantly be improved.
In this section, the status and prospects for several benchmark
rare decays of different strange-hadron species are shown.

\subsection{\boldmath Rare decays of \ks mesons}

Due to its shorter lifetime compared to \KL and \Kp, the \KS meson is the most accessible in terms of reconstruction in LHCb. With a geometric acceptance at the $1\%$ level and a production
cross section of about 0.3 barn, the LHCb Phase-II upgrade could reach branching fraction sensitivities
down to the level of $10^{-15}$ in the ideal case of perfect selection and trigger with no background.
In the following, the channels LHCb has already investigated  are discussed in addition to new analysis suggestions.

\subsubsection{\boldmath \Ksmm}

\begin{figure}[t!]
\centering
\includegraphics[width=0.8\textwidth]{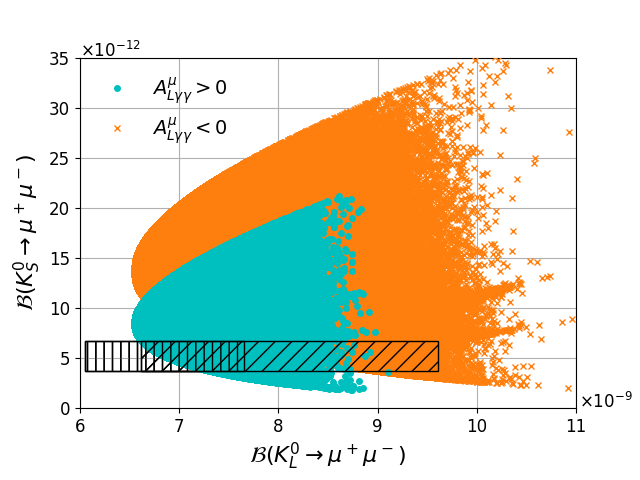}
\caption{\label{fig:KmmMSSM} A generic scan of $\mathcal{B}(K_S^0\rightarrow\mu^+\mu^-)$ {\it vs} $\mathcal{B}(K_L^0\rightarrow\mu^+\mu^-)$ from Ref.~\cite{Chobanova:2017rkj}, in an MSSM scenario with
universal gaugino masses. The cyan dots correspond to predictions with $A^\mu_{L\gamma \gamma} > 0$ and the orange crosses to 
predictions using $A^\mu_{L\gamma \gamma} < 0$. The vertically hatched area corresponds to the SM prediction for $A^\mu_{L\gamma \gamma} > 0$ 
while the diagonally hatched area corresponds to the SM prediction for $A^\mu_{L \gamma \gamma} < 0$. }
\end{figure}

In the SM, the \Ksmm decay is dominated by long-distance (LD) effects with subdominant short-distance (SD) contributions coming from
$Z$-penguin and $W$-box diagrams. Yet in absolute terms, the long-distance contribution is still minute with the decay
rate highly suppressed~\cite{Ecker:1991ru,Isidori:2003ts,DAmbrosio:2017klp}. The theoretical prediction,
\begin{equation*}
  \BRof\Ksmm_{\rm SM}  = (5.18\pm1.50_{\rm LD}\pm 0.02_{\rm SD})\times 10^{-12}\quad ,
\end{equation*}
\noindent when compared with the current experimental upper limit~\cite{LHCb-PAPER-2017-009}
\begin{equation*}
  \BRof\Ksmm < 8 \times 10^{-10} ~{\rm{at}}~90\%~{\rm{CL}}\quad ,
\end{equation*}
leaves room for small BSM contributions to interfere and compete with the SM rate. This is shown
to be the case in leptoquark (LQ) models~\cite{Dorsner:2011ai,Bobeth:2017ecx} as well as in the Minimal Supersymmetric Standard Model (MSSM)~\cite{Chobanova:2017rkj}.
In the LQ case, the enhancements can reach as high as the current experimental bound, while within the MSSM, \BRof\Ksmm can adopt values anywhere in the range $[0.78, 35.00]\times 10^{-12}$, even surpassing the 
experimental bound in certain narrow, finely-tuned regions of the parameter space~\cite{Chobanova:2017rkj}. This can be seen in figure~\ref{fig:KmmMSSM}, where $A^\mu_{L\gamma \gamma}$ indicates
the long-distance contribution to \BRof\Klmm. The \CP asymmetry of this decay
is also sensitive to BSM contributions, but experimentally accessible only by means of a tagged analysis.

\begin{figure}[t!]
\centering
\includegraphics[width=0.8\textwidth]{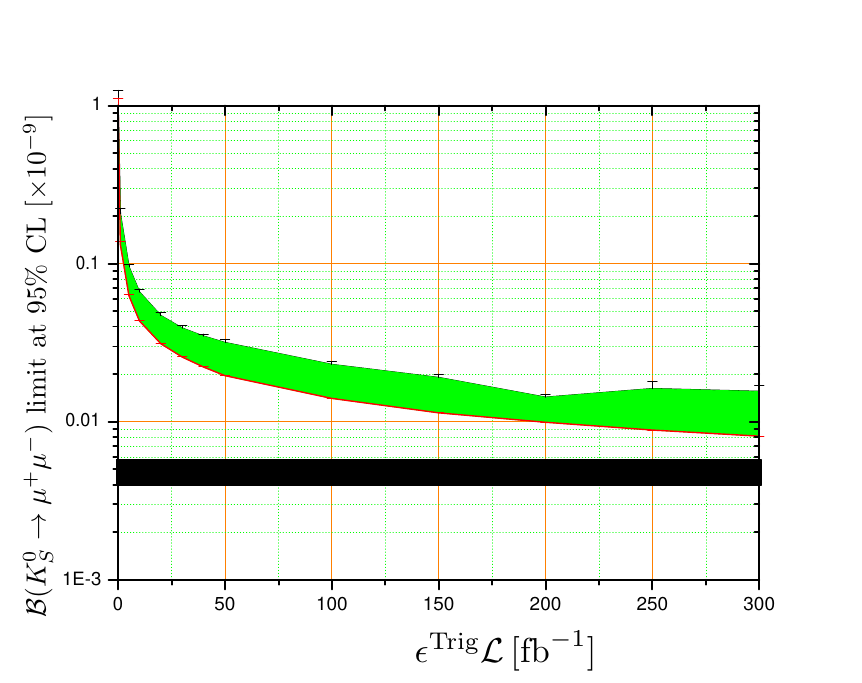}
\caption{\label{fig:KmmProspects} Expected upper limit of \BRof\Ksmm from LHCb including upgrades, against the product of the integrated luminosity and trigger efficiency, taken from Ref.~\cite{Santos:2018zbz}. The LHCb upgrade is expected to collect 50\invfb, and the Phase-II $\approx 300\invfb$. 
}
\end{figure}

The LHCb prospects for the search for \Ksmm decays are excellent. With only 2011 data, the experiment improved the previous world upper limit
by a factor of thirty~\cite{LHCb-PAPER-2012-023} and recently gained another factor of ten~\cite{LHCb-PAPER-2017-009}. In the case of an LHCb Phase-II upgrade running during the proposed HL-LHC era, the full software trigger will allow an exploration of branching fractions below the $10^{-11}$ regime. 
Figure~\ref{fig:KmmProspects}, first shown in Ref.~\cite{Santos:2018zbz}, shows the expected upper limit
of \BRof\Ksmm as a function of the integrated luminosity scaled by the trigger efficiency,
  based on the extrapolation given in Ref.~\cite{LHCb-PAPER-2017-009}. 
This demonstrates that if the trigger efficiency is near $\sim 1$, 
as can be achieved technically with the Phase-I full software trigger, LHCb could exclude branching fractions down towards the vicinity of the SM prediction.

\subsubsection{\boldmath \Kspizmm}

The experimental uncertainty on \BRof\Kspizmm is the dominant uncertainty on the SM prediction of \BRof\Klpizmm, the latter being an important
channel for BSM searches, such as extra dimensions~\cite{Bauer:2009cf}. Currently, the only existing measurement comes from the NA48 experiment~\cite{Batley:2004wg},
\begin{equation*}
\BRof\Kspizmm = (2.9_{-1.2}^{+1.5}\pm0.2)\times 10^{-9}\quad .
\end{equation*}
The upgraded LHCb experiment can quickly eclipse NA48 in terms of precision on \BRof\Kspizmm and achieve a level of $0.25\times 10^{-9}$ with 50~\invfb of integrated luminosity and assuming 100\% trigger efficiency~\cite{Chobanova:2195218} 
(see footnote~\ref{footnote:trigger}). 

Aside from the branching fraction, the differential decay rate in the dimuon mass possesses interesting information. As the electromagnetic structure of this decay in the SM receives only a 
single contribution from the vector current, an amplitude analysis cannot offer any advantages over a fit to the dimuon mass spectrum alone. The decay dynamics of this channel are assumed to 
be governed by a linear dependence in $q^2$, thus there are two free, real parameters of the model, which can be determined from data, $a_S$ and $b_S$, where $b_S$ is the coefficient of the linear 
term in $q^2$. 
This complements the information available from the branching fraction, which has the form,
\begin{equation*}
  \BRof\Kspizmm \propto 0.07 - 4.52a_S - 1.5b_S + 98.7a_S^2 + 57.7 a_Sb_S + 8.95b_S^2\quad ,
\end{equation*}
in the SM~\cite{DAmbrosio:1998gur}.

Importantly, $a_S$ is the relevant parameter for the SM determination of \BRof\Klpizmm. It has been estimated from the NA48 measurement of \BRof\Kspizmm that  $|a_S| = 1.2\pm 0.2$~\cite{Bauer:2009cf}, assuming
vector meson dominance (VMD), where $b_S/a_S = m^2_K/m^2_\rho$. Without VMD,
resolving $a_S$ with only a single observable is not possible.
Hence, as the precision in \BRof\Kspizmm increases, use of the $q^2$ dependence, which has been calculated in Ref.~\cite{DAmbrosio:1998gur}, becomes a viable approach in avoiding this model dependence.

Two degenerate solutions are expected for both $a_S$ and $b_S$. A pseudo-experiment study indicates that the significance of the 
sign-flip in $a_S$ is consistent with zero even up to signal yields well beyond the reach of any proposed LHC upgrade. Although the model-dependent expectation is that the product $a_Sb_S$, is 
preferred to be positive, the proximity to zero of the $b_S$ solution corresponding to negative $a_S$ renders this constraint untenable.

A number of analysis configurations from a purely statistical point of view are considered, neglecting systematic uncertainties. 
The statistical power has been obtained from the expected sensitivity in 
\BRof\Kspizmm, where the signal plus background yield is translated into an effective signal-only yield. 
Firstly, the scenario where both $a_S$ and $b_S$ are measured from the $q^2$ 
distribution is considered. An additional constraint coming from NA48 is also considered, which relates the branching fraction of \Kspizee, to $a_S$ and $b_S$ through
\begin{equation*}
  \BRof\Kspizee = [0.01 - 0.76a_S - 0.21b_S + 46.5a_S^2 + 12.9 a_Sb_S + 1.44b_S^2] \times 10^{-10}\quad .
\end{equation*}
The uncertainty on $a_S$ using the value of $b_S$ motivated by VMD is also investigated. In this paradigm, it becomes possible to measure $a_S$ from the \Kspizmm yield alone, which is tested as the final case.

\begin{table}[h]
  \footnotesize
  \centering
  \caption{Projected statistical uncertainties on $a_S$ under various analysis conditions.}
  \label{tab:kspi0mumu}
  \begin{tabular}
    {@{\hspace{0.5cm}}c@{\hspace{0.25cm}}  @{\hspace{0.25cm}}c@{\hspace{0.25cm}}  @{\hspace{0.25cm}}c@{\hspace{0.5cm}}}
      \toprule
    Configuration & ~~~Phase I~~~ & ~~~Phase II~~~ \\ \midrule
    BR \& $q^2$ fit & 0.25 & 0.10\\ 
    BR \& $q^2$ fit with NA48 constraint & 0.19 & 0.10\\ 
    BR \& $q^2$ fit fixing $b_S$ & 0.06 & 0.024\\
    $a_S$ measurement from BR alone & 0.06 & 0.024\\
  \bottomrule
  \end{tabular}
\end{table}

The reach of LHCb in each of these scenarios is summarised in table~\ref{tab:kspi0mumu} for different effective yields. In the case that $b_S$ is measured from the data, its uncertainties are 
expected to be 0.87 (0.35) for the Phase-I (Phase-II) data samples. The results show that with the effective events from Phase-I data, the constraint coming from NA48 on the \Kspizee branching fraction will play a 
role in reducing the uncertainty on $a_S$, while with Phase-II data, the uncertainty will be entirely dominated by the LHCb \Kspizmm measurement.
The results also indicate the vast improvement in $a_S$ that 
becomes possible at the expense of model independence and demonstrate that the $q^2$ distribution has very little impact on the overall uncertainty on $a_S$ when $b_S$ is fixed. Further improvements could, of course,  come from an LHCb
measurement of \Kspizee.

\subsubsection{\boldmath \Kspipiee and other \KS dielectron modes}

With a relatively high branching fraction of $\sim 5 \times 10^{-5}$ \cite{Batley:2011zza}, the \Kspipiee decay offers an excellent opportunity to study rare decays of \KS mesons to electrons at LHCb. 
Due to bremsstrahlung, electrons are generally more difficult to reconstruct than other particles, such as pions or muons. This is especially the case for low momentum electrons, such as those expected in \KS decays. Given the branching fraction of \Kspipiee, a significant yield per \invfb is expected to be produced within the LHCb acceptance, thus this decay could be used both for \CP-violation studies~\cite{Batley:2011zza} and to search for potential resonant structure in the $e^+e^-$ invariant-mass spectrum. From a purely experimental standpoint, it is interesting for the study of both the reconstruction and identification of low momentum electrons and to harness as a normalisation channel for various 4-body \KS rare decays. Examples include decays to four leptons, which could be sensitive to the presence of BSM contributions \cite{DAmbrosio:2013qmd}, suppressed SM decays such as $\KS \to \pi^+\pi^-\mu^+\mu^-$, or Lepton Flavour Violating decays like $\KS \to \mu^+\mu^+e^-e^-$ and $\KS \to \pi^+\pi^-\mu^+e^-$. Moreover, \Kspipiee could present as a prominent background in these searches, ergo, a comprehensive understanding of its expected yield and invariant-mass distribution becomes crucial.

\begin{figure}[t!]
\centering
\includegraphics[width=0.65\textwidth]{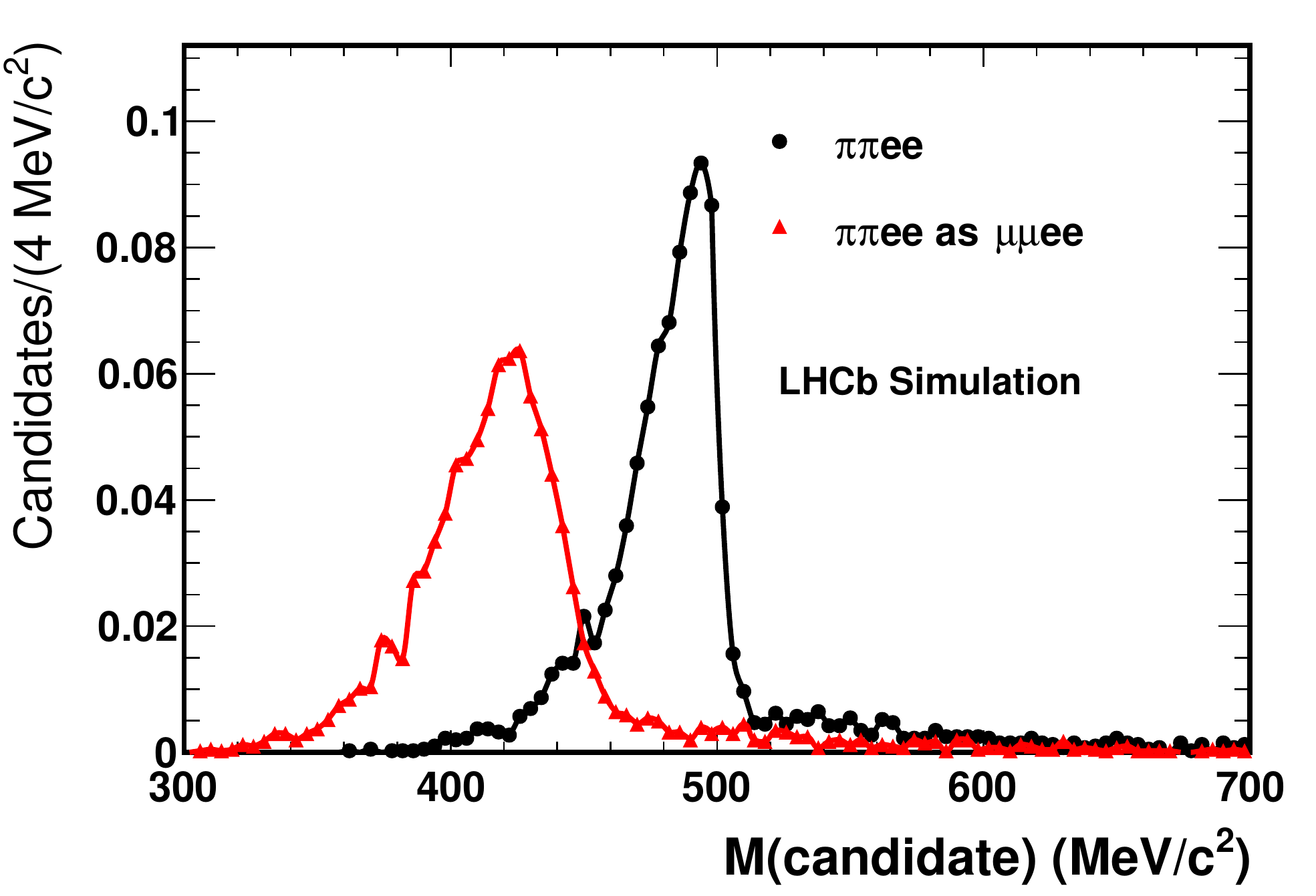}
\caption{\label{fig:kspipiee} Expected invariant-mass shape of \Kspipiee shown additionally with the $\mu^+\mu^-e^+e^-$ mass hypothesis, taken from Ref.~\cite{MarinBenito:2193358}.}
\end{figure}

The \Kspipiee decay at LHCb is studied in Ref.~\cite{MarinBenito:2193358}. This analysis involves a generic study of the decay using LHCb simulated samples and includes a search with the Run~1 data, giving prospects for Run 2 and Run 3. The LHCb hardware trigger is found to limit observation of this decay, with only $\sim$100 candidates per \invfb expected to be reconstructed and selected in Run 1 and Run 2. Despite this relatively low yield, it is also concluded that a purpose-built offline selection, including the use of a Multi-Variate Analysis (MVA) classifier, could lead to an observation of the signal. The prospects for Run 3 are much better, with an expected yield at the level of $\sim 50 \times 10^{3}$ selected candidates per \invfb. Furthermore, the presence of \Kspipiee as a background for 4-lepton final states is also studied. Figure~\ref{fig:kspipiee}, taken from Ref.~\cite{MarinBenito:2193358}, shows the invariant-mass shape of the \Kspipiee decay in conjunction with the alternate $\mu^+\mu^-e^+e^-$ mass hypothesis, to highlight its separation with respect to a potential $\KS \to \mu^+\mu^-e^+e^-$ signal, both obtained from simulation. While both peaks are separated, a significant contamination from \Kspipiee is expected in the signal region due to the long tails of the distribution and the much larger yield expected for this mode. However this contribution can be modelled from simulation and systematic effects controlled with data, in analogy to the contamination of \Kspipi decays as a background for \Ksmm~\cite{LHCb-PAPER-2017-009}.

The presence of electron bremsstrahlung combined with the low transverse momentum of the final state particles, makes the invariant-mass resolution of this final state significantly worse when compared to \Ksmm, for instance. New reconstruction strategies could enhance the sensitivity of LHCb to \Kspipiee and other similar final states, such as those mentioned above. Given that the position of the \KS production and decay vertices can be determined, the invariant-mass resolution of the \KS could be  
calculated ignoring the absolute momentum of one of the four final state particles through relativistic kinematic constraints. This is advantageous as the invariant-mass resolution becomes less dependent on bremsstrahlung, given that the direction of electrons in the VELO is barely influenced by such effects. In addition, this technique could allow a more efficient reconstruction of these electrons, using tracks not required to have a segment after the magnet. Taking into account that the VELO pattern recognition efficiency is at the level of $\sim 70\%$ \cite{Hutchcroft:1023540}, even for tracks with $p\sim {\cal O}(1 \mevc)$, improvements in the reconstruction efficiency up to a factor of 10 could be theoretically possible.

\subsubsection{\boldmath \Ksgmm, $\KS\to X^0\mu^+\mu^-$ and $\KS\to X^0 \pi^{\pm}\mu^{\mp}$}
\begin{figure}[t!]
  \includegraphics[width=0.49\textwidth]{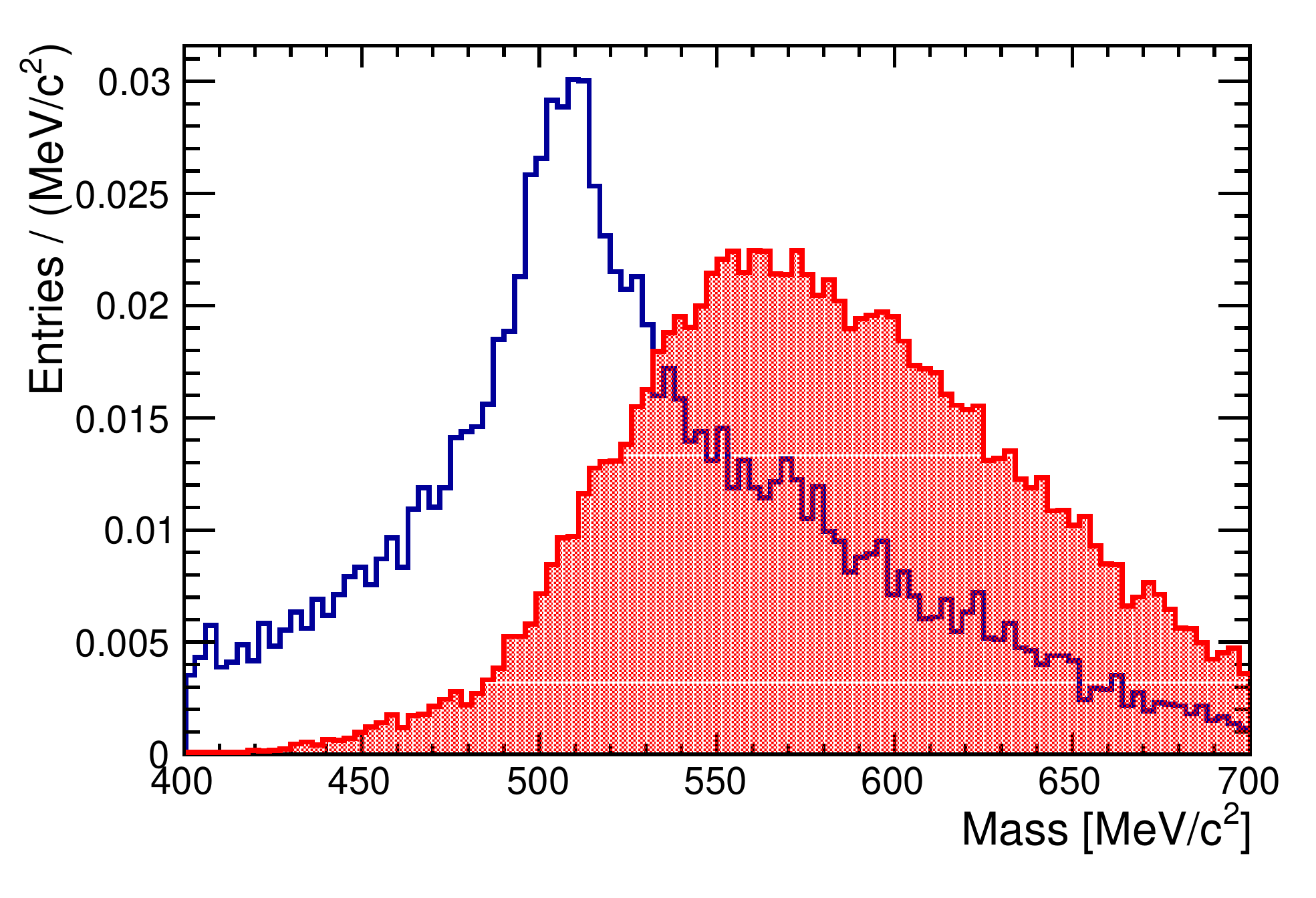}
  \includegraphics[width=0.49\textwidth]{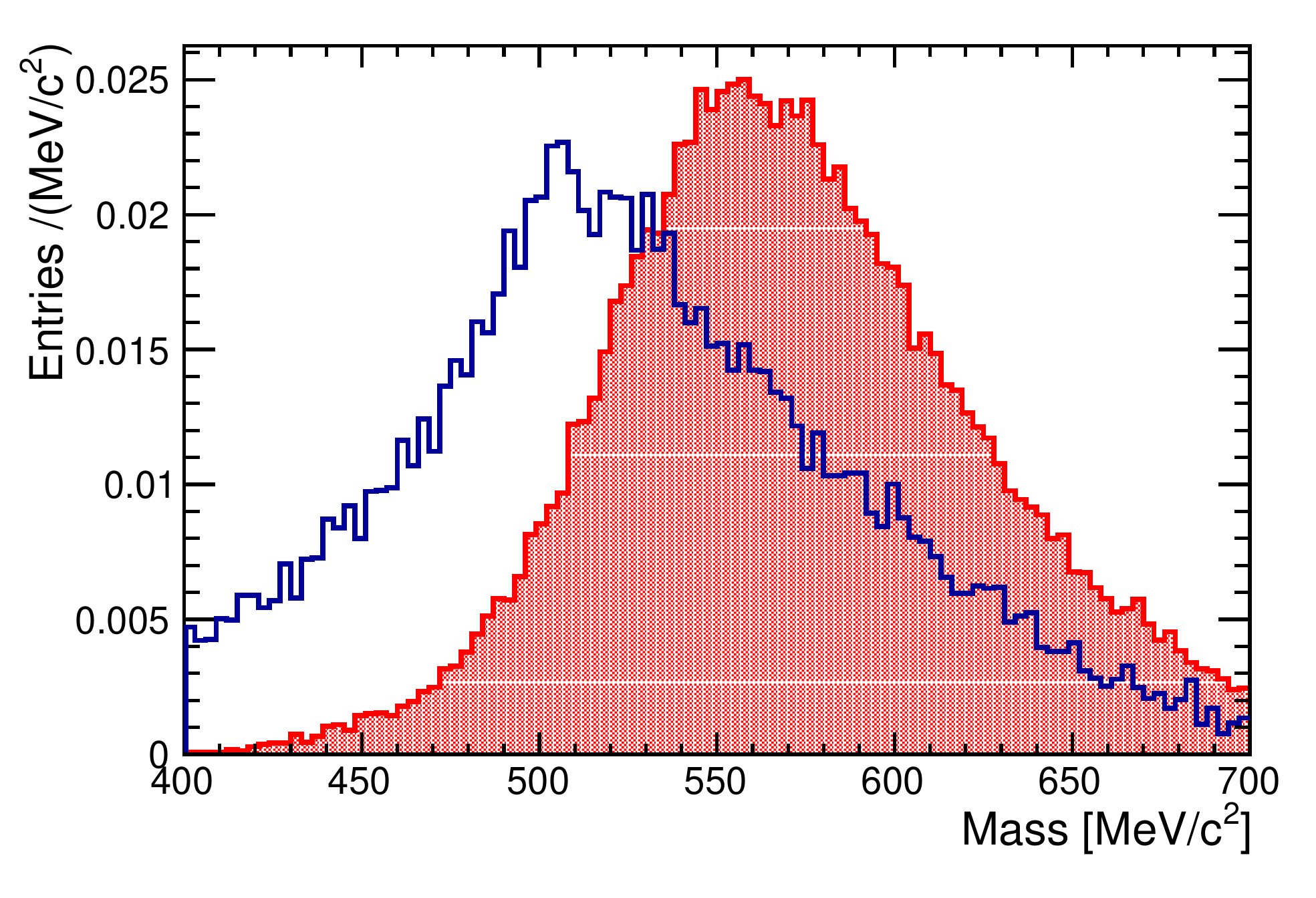}
  \includegraphics[width=0.49\textwidth]{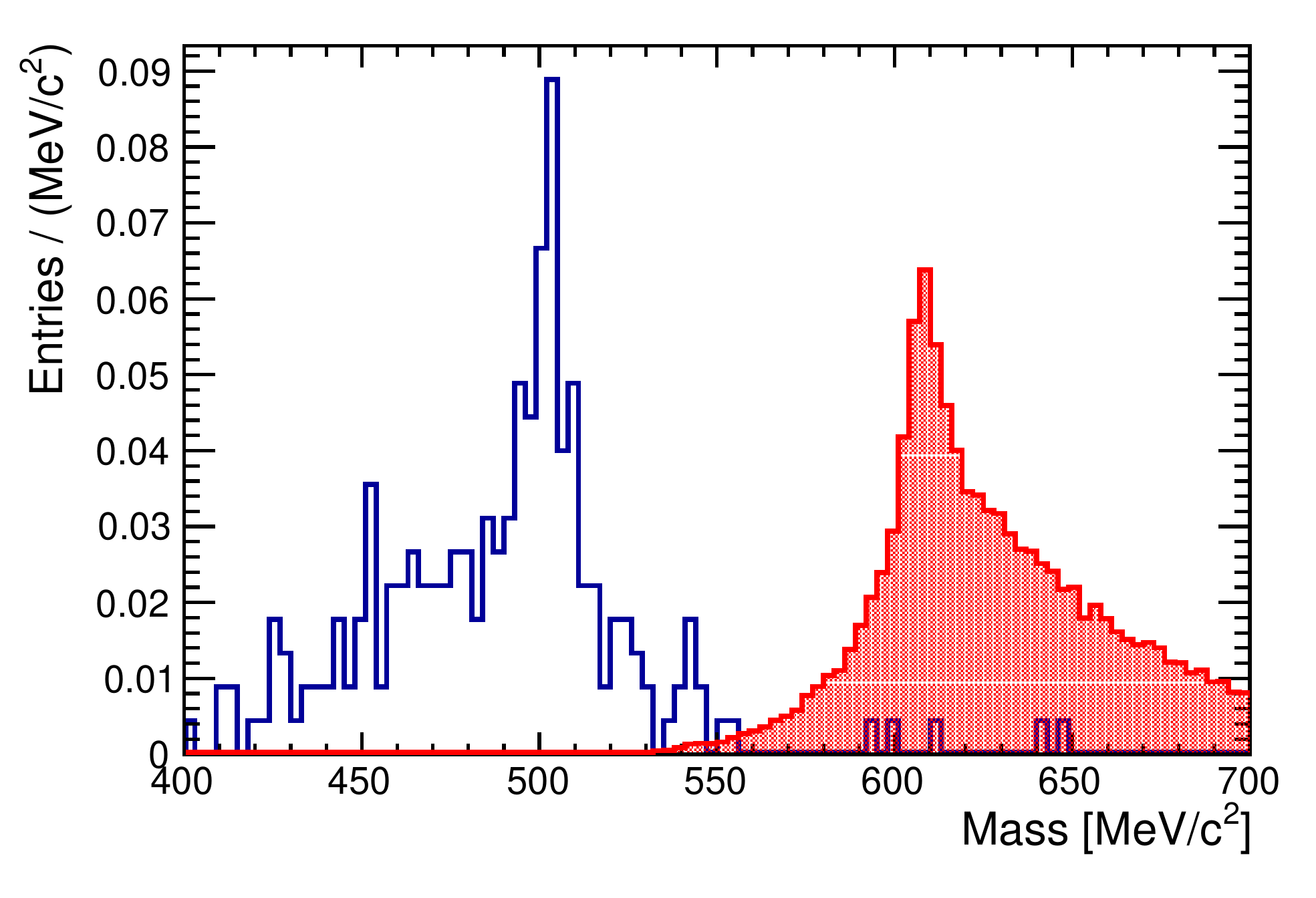}
  \includegraphics[width=0.49\textwidth]{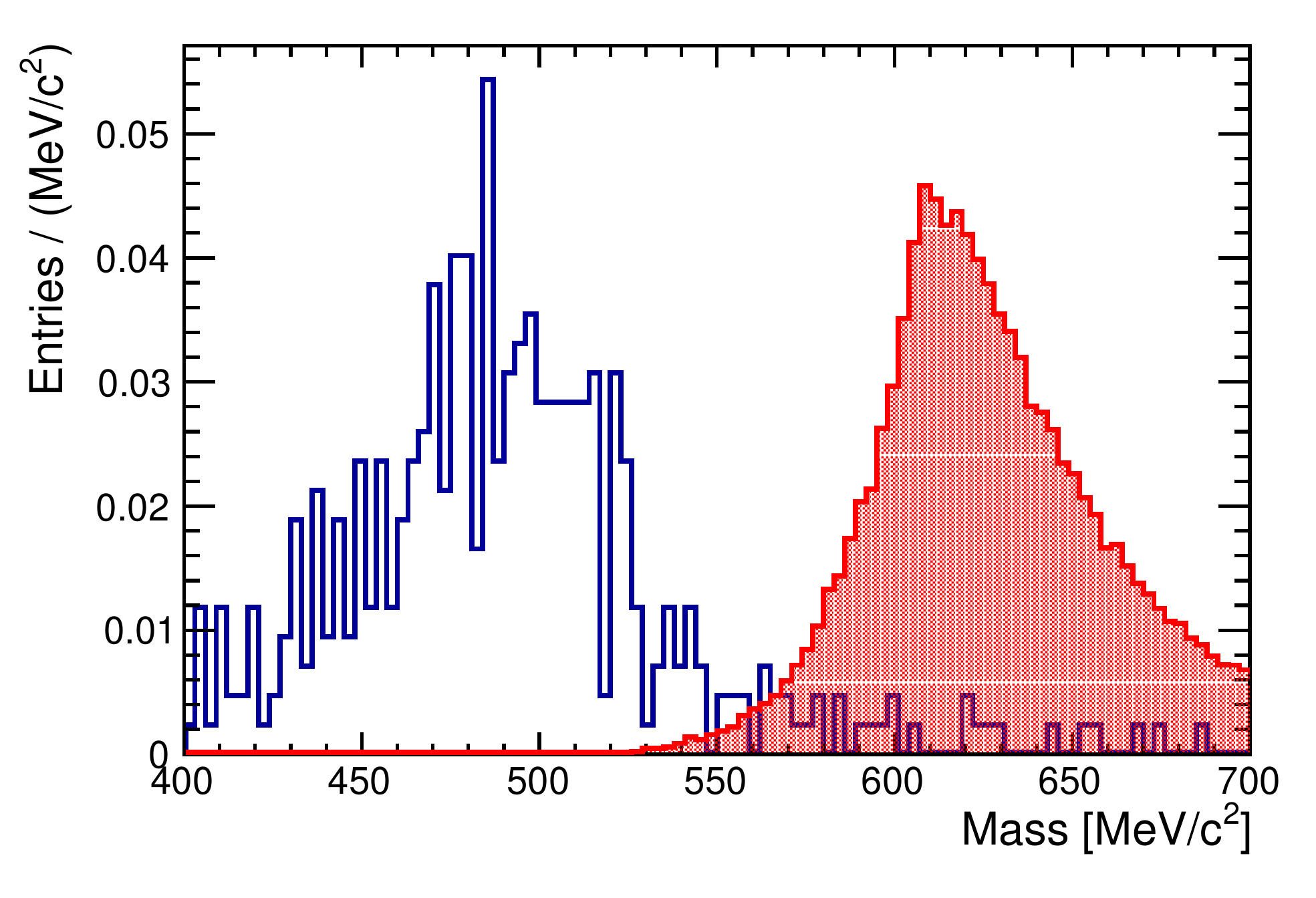}
 
  \caption{\label{fig:mmg} Reconstructed invariant mass for \Ksgmm (top) and \Kspizmm (bottom) obtained from simulation. The \Ksgmm and \Kspizmm signal events are shown
    with a solid blue line and the \Kspipi background is illustrated with  red filled histograms. The left side portrays events reconstructed with long tracks, while reconstruction with downstream tracks are depicted on the right.}
\end{figure}

The analysis strategy of \Kspizmm can be applied to any $\KS\to X^0\mu^+\mu^-$ mode, where $X^0$ is an arbitrary neutral system. The performance of the search will be strongly related to the mass
of the neutral system, with heavier $X^0$ leading to superior invariant-mass resolution of the \KS peak. 
The resolution is studied here using simulated \Ksgmm decays, corresponding to the most restrictive case of a massless $X^0$.
 This decay is predicted in the SM to occur with a branching fraction of $(1.45\pm0.27)\times10^{-9}$ ~\cite{Colangelo:2016ruc}.
 Background from generated \Kspipi is also considered with the aforementioned simplified tracking emulation. From figure~\ref{fig:mmg}, the distinction between signal and background is visibly worse for \Ksgmm than it is for \Kspizmm. Nevertheless, both peaks
show clear separation and hence the search is feasible. A reduction of the \Kspipi background is possible by requiring the dimuon candidate to point away from the primary vertex, in the same way as is done in \Kspizmm analysis~\cite{Chobanova:2195218}.
A similar strategy can be embraced in $\KS\to X^0 \pi^{\pm}\mu^{\mp}$, where the $X^0$ in this case could be some neutrino, either from the SM decay $\KS\to \pi^{\pm}\mu^{\mp} \nu$ or a heavy BSM neutrino (see also section~\ref{sec:semiks}).

\subsection{\boldmath Rare decays of \kplus mesons}

From the efficiency ratios of table~\ref{tab:acc} and considering that
sensitivities for \KS branching fractions are at the $10^{-10}$-$10^{-12}$ level, sensitivities from $10^{-7}$-$10^{-10}$ could be expected for \Kp decays, depending on the background level.
For \kplus mesons, which are electrically charged and long-lived, the possibility to interact with one or more VELO stations can lead to an additional source of discrimination against combinatorial background~\cite{Contu:1693666}.
Single event sensitivities could then well reach below $10^{-12}$, in the case of very small background (muonic channels), 
while taking into account higher levels of background, possible sensitivities of order $10^{-10}-10^{-11}$ are foreseen. 

\subsubsection{\boldmath \Kppimumu and \kpiee}

The decays $K^\pm \rightarrow \pi^\pm \mu^+ \mu^-$  are flavour-changing processes induced at the one-loop level, which are well suited to explore SM structure and its extensions.  These decays are dominated by long-distance contributions involving one photon exchange {\it i.e}. $K \rightarrow \pi \gamma^* \rightarrow \pi \mu^+ \mu^-$. 
The branching fraction has been derived within the framework of Chiral Perturbation Theory ($\chi_{PT}$) in terms of a vector-interaction form factor, which describes the single-photon exchange and characterises the dimuon invariant-mass spectrum~\cite{DAmbrosio:1998gur,Friot:2005pe,Dubnickova:2006mk}.  
The differential decay rate can be written as a kinematic term depending on masses and 4-momenta, multiplied by $|W(z)|^2$, where $W$ is the form factor and $z= ( m_{\mu\mu} / M_K )^2 $.
The form factor is given by $W(z) \propto  W_{\rm pol} (z) W_{\pi\pi} (z) $, where the second term represents the tiny contribution from the two-pion-loop intermediate state and the first term is phenomenologically described by a polynomial. As the form factor is required to vanish at lowest order in the low-energy chiral expansion, the polynomial term takes the form $W_{\rm pol} (z) =  ( a_+ + b_+ z ) $, 
where $a_+$ and $b_+$ are free parameters of the model to be determined by experiment.
In a similar fashion to $b \rightarrow s$ transitions, $s \rightarrow d$ processes can be described with an effective Lagrangian depending on Wilson coefficients, generating only the non-zero Wilson coefficients $C_{7A}$ and $C_{7V}$ for the semileptonic operators.
Such coefficients can be split into SM and BSM contributions. In particular, $a_+$ can be written 
as a function of the Wilson coefficient $C_{7A}$~\cite{Crivellin:2016vjc}, leading to potential constraints on BSM. A further comparison of the electron and muon channels 
would provide an additional test of Lepton Flavour Universality and further constrain BSM dynamics. 

Natural extensions of the SM involve the inclusion of sterile neutrinos which mix with ordinary neutrinos. An example is the Neutrino Minimal Standard Model ($\nu$MSM)~\cite{Asaka:2005pn}, which can be further extended by adding a scalar field to incorporate inflation and provide a common source for electroweak symmetry breaking and right-handed neutrino masses~\cite{Shaposhnikov:2006xi}. The new particles predicted by these models can be produced in charged kaon decays. Notably, the two-unit Lepton Number Violating (LNV) $K^\pm  \rightarrow \pi^\mp \mu^\pm \mu^\pm$ decay could proceed via an off-shell or on-shell Majorana neutrino~\cite{Littenberg:2000fg}, while an inflaton could be produced in the Lepton Number Conserving (LNC) $K^\pm \rightarrow \pi^\pm X$, decaying promptly to $X \rightarrow \mu^+ \mu^-$~\cite{Bezrukov:2009yw}.

\begin{figure}[t!]
  \includegraphics[width=0.49\textwidth]{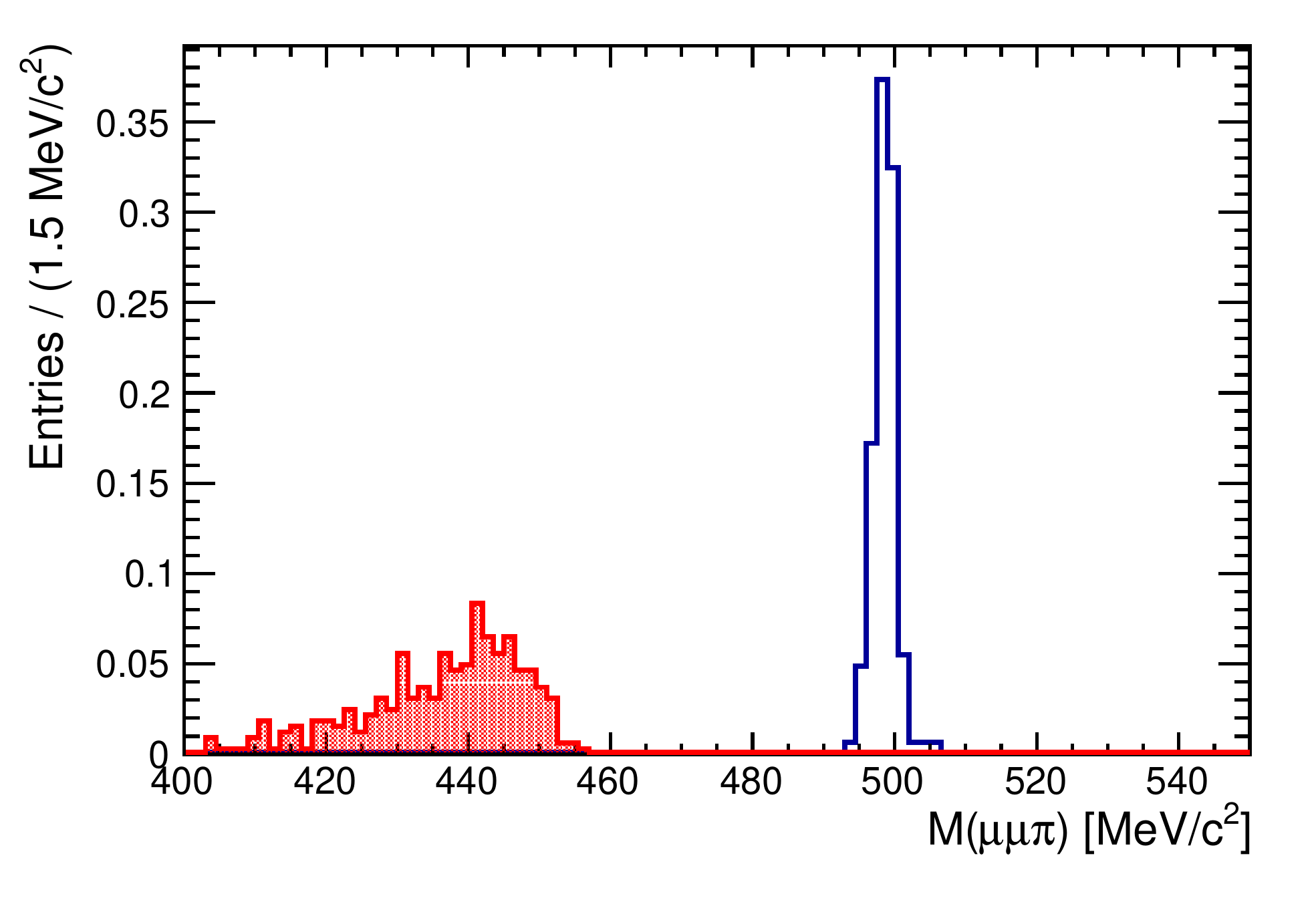}
  \includegraphics[width=0.49\textwidth]{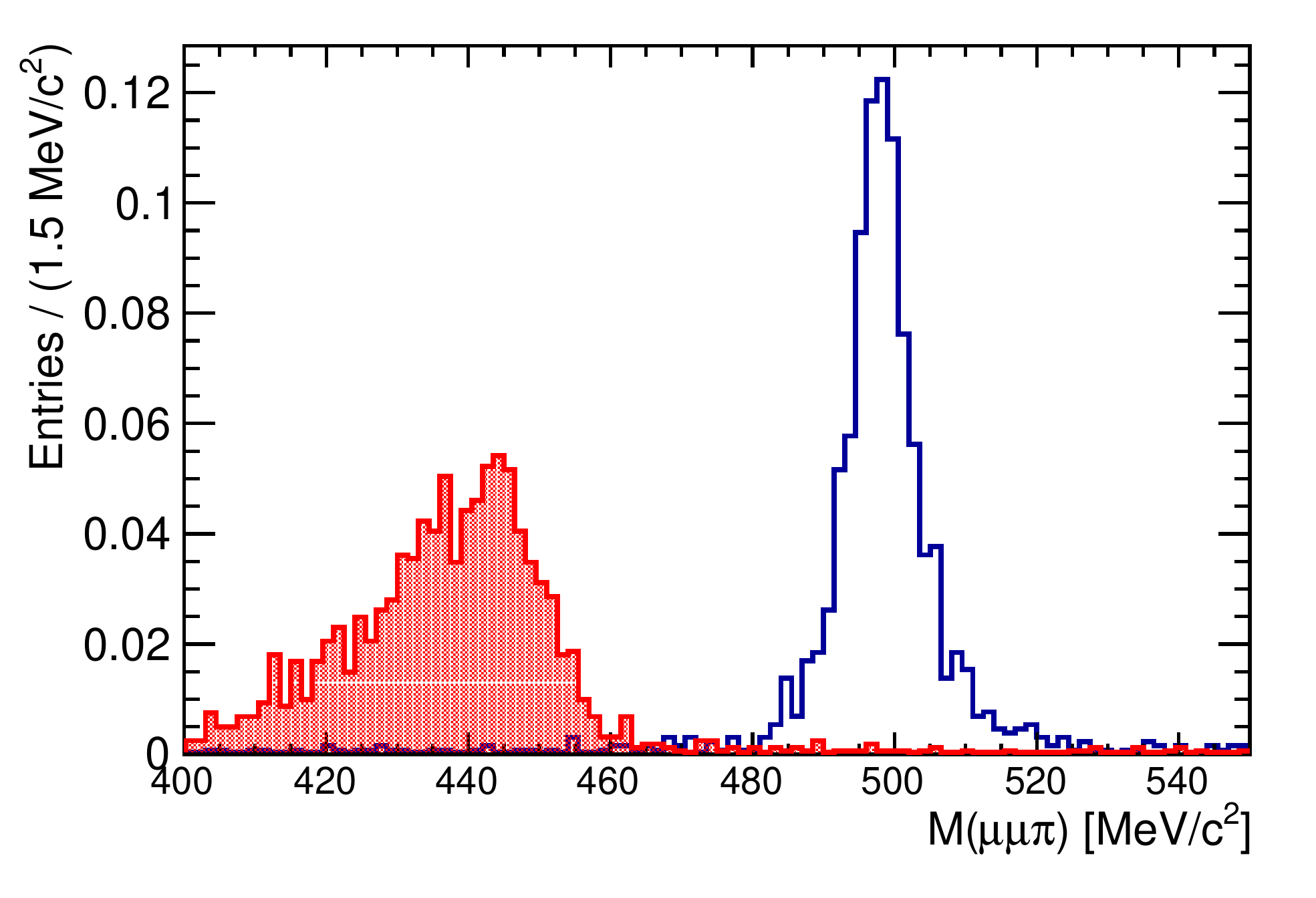}

  \caption{\label{fig:Kpimm} Reconstructed invariant mass for \Kppimumu, where signal events are shown
    with a solid blue line and \KpTpi background  illustrated by red filled histograms. The left side gives events reconstructed with long tracks, while reconstruction with downstream tracks are pictured on the right.}
\end{figure}

The NA48/2 collaboration~\cite{Batley:2239327,Batley:2011zz} reports the most precise measurement to date of the branching fraction and provide limits on the Majorana neutrino and inflaton. 
They measured
$${\cal B} (K^\pm \rightarrow \pi^\pm \mu^+ \mu^- ) = (9.62 \pm 0.21_{\rm stat} \pm 0.13_{\rm syst} ) \times10^{-8},  $$
$${\cal B}( K^\pm  \rightarrow \pi^\mp \mu^\pm \mu^\pm ) < 8.6 \times 10^{-11} \,\,\, (90\%\; {\rm CL}), $$
$${\cal B}( K^\pm \rightarrow \pi^\pm X ) < 10^{-11} {\text -} 10^{-9} \,\,\, (90\%\; {\rm CL}), $$
where the range depends on the assumed resonance lifetime.
The NA62 experiment plans to improve on all these measurements and limits, though with positively-charged kaons only~\cite{NA62:2312430}.
The LHCb mass resolution is sufficient to separate these decays from the kinematically similar \KpTpi, as illustrated in figure~\ref{fig:Kpimm}.
LHCb can acquire large \Kppimumu signal yields as table~\ref{tab:acc} and figure~\ref{fig:minbias} clearly indicate. Assuming ${\cal O}(1)$ trigger efficiencies, a yield of ${\cal O}$(10$^4$) fully reconstructed
and selected signal events is expected per year of upgraded-LHCb data taking, even considering only long-track candidates. 
This suggests \Kppimumu decays would provide an early opportunity
for a measurement to demonstrate the potential of the upgraded detector for these channels. 
Similar arguments apply to the \kpiee mode, whose somewhat lower reconstruction efficiency due
to the presence of electrons is negated by its larger branching fraction. Rigorous control over the systematic uncertainties will be paramount 
in order to improve the current world-average precision of $3\%$ on the electron mode. If successful, the full spectrum of both channels  will afford a highly precise test of Lepton Flavour Universality.

\subsection{Tests of LFV} \label{sec:LFV}

Modes with LFV, such as $K \to (n\pi) \mu^\pm e^\mp$ form null tests of the SM. Sizeable BSM contributions to such decays have garnered increased attention in recent times because of hints at Lepton Universality Violation (LUV) in $B \to K^{(\ast)} \ell^\pm \ell^\mp$ processes. In fact, both classes of processes can be generated by new contributions to the product of two neutral currents, involving down-type quarks and leptons respectively, the only difference being the strength of the flavour couplings involved.

From the amount of LUV alluded to in $B \to K^{(\ast)} \ell^\pm \ell^\mp$, one may expect $B \to K^{(\ast)}$ LFV rates of the order of $10^{-8}$ using general effective-theory (EFT) arguments~\cite{Glashow:2014iga}. More quantitative estimates require the introduction of a flavour model \cite{Guadagnoli:2015nra,Boucenna:2015raa,Celis:2015ara,Alonso:2015sja,Gripaios:2015gra,Crivellin:2016vjc,Becirevic:2016zri,Becirevic:2016oho,Hiller:2016kry,Becirevic:2017jtw,Buttazzo:2017ixm,King:2017anf,Bordone:2018nbg}.
As discussed in Ref.~\cite{Borsato:2018tcz}, such arguments can be extended to the $K \to (\pi) \mu^\pm e^\mp$ case, with fairly general assumptions on the different flavour couplings involved. Expected rates can be as large as $10^{-10}$ - $10^{-13}$ for the $K_L \to \mu^\pm e^\mp$ mode and a factor of $\sim 100$ smaller for $K^+ \to \pi^+ \mu^\pm e^\mp$. Taking into account the suppression mechanisms at play, such `large' rates are a non-trivial finding. Their relatively wide range is due to the inherent model dependence especially in the choice of the leptonic coupling and the overall scale of the new interaction, typically between $5$ and $15$ TeV~\cite{Borsato:2018tcz}. 
Since limits on the branching fractions for the $K \to \pi e \mu$ modes were pushed down to the level of $10^{-11} - 10^{-12}$ in the 1990s, 
there has been no significant further progress on the experimental side, with the current limits at 90\% CL,
\begin{eqnarray}
\label{eq:limits}
\begin{tabular}{clcl}
$\mathcal B(K_L \to e^\pm \mu^\mp) < 4.7 \times 10^{-12}$ & \hspace{-0.3cm}\cite{Ambrose:1998us}~, &~~~~
$\mathcal B(K_L \to \pi^0 e^\pm \mu^\mp) < 7.6 \times 10^{-11}$ & \hspace{-0.3cm}\cite{Abouzaid:2007aa}~,\\
[0.1cm]
$\mathcal B(K^+ \to \pi^+ e^- \mu^+) < 1.3 \times 10^{-11}$ & \hspace{-0.3cm}\cite{Sher:2005sp}~, &~~~~
$\mathcal B(K^+ \to \pi^+ e^+ \mu^-) < 5.2 \times 10^{-10}$ & \hspace{-0.3cm}\cite{Appel:2000tc}~,\\
\end{tabular}
\end{eqnarray}
being decades old.

 These modes can be profitably pursued at the upgraded LHCb, benefiting from huge strange-production yields. In fact, starting from a total $K^\pm$ cross section of 0.63~barns and taking into account the fraction of kaons in the pseudorapidity acceptance of LHCb, one can estimate a $K^\pm$ cross section as large as $0.14$ barns. 
Ref.~\cite{Borsato:2018tcz} presents a feasibility study of the modes listed in eq.~\eqref{eq:limits}, taking $K^+ \to  \pi^+ \mu^\pm e^\mp$ as a benchmark.
It can be seen that LHCb may be able to update the existing limits and probe a sizeable part of the parameter space suggested by the discrepancies in $B$ physics.

\subsection{\boldmath Rare decays of $\Sigma$ hyperons}

LHCb has recently published the most precise search for \sigmapmumu~\cite{LHCb-PAPER-2017-049}, showing strong evidence 
for this decay with $4.1\sigma$ significance. A measurement of the branching fraction is reported  along with a dimuon invariant-mass distribution consistent with SM predictions, challenging the so-called HyperCP-anomaly~\cite{Park:2005eka}. 
This measurement was based on Run 1 data, where no trigger path existed specifically for this channel. 
As discussed in Ref.~\cite{Dettori:2297352}, Run 2 will have a dedicated trigger both at the HLT1 and HLT2 levels, where about an order of magnitude increase in the trigger efficiency is anticipated. With a signal yield in excess of 150 events, Run 2 data
will allow a measurement of the differential decay rate and possibly 
other observables with recent predictions such as the forward-backward asymmetry~\cite{He:2018yzu}. 
Applying similar reasoning on the trigger efficiency as with other decays in this document, on the order of a thousand signal decays could be measured per year of data taking with an upgraded LHCb detector, opening the possibility for precision measurements of direct \CP violation.
Assuming similar reconstruction and selection efficiencies, a search for the lepton and baryon number violating  \sigmapmumulfv decay
could also be performed, reaching an expected branching fraction sensitivity on the order of $10^{-9}$. 

While of great interest, it will be difficult for LHCb to improve the precision 
on the branching fraction of the radiative \sigmapgamma decay, whose world average is currently $\mathcal{B}(\sigmapgamma ) = (1.23\pm 0.05)\times 10^{-3}$~\cite{PDG}. 
On the other hand, the ability to 
reconstruct the \sigmappiz decay, which has similar topology in the detector, has already been demonstrated~\cite{LHCb-PAPER-2017-049}.
This implies that the \sigmapgamma decay could be useful as an alternative normalisation channel, 
particularly in a possible search for \sigmapee decays. 
By virtue of the electron mass, this channel
receives a larger contribution from long-distance photon contributions compared to \sigmapmumu, for a predicted branching fraction of $\mathcal{B}(\sigmapee) \in [9.1,10.1] \times 10^{-6}$~\cite{He:2005yn}. 
The only experimental information available on this channel dates back to 1969 where three events where observed leading to an upper limit of $7 \times 10^{-6}$ at 90\% CL~\cite{Ang:1969hg}.
Unsurprisingly, this yield is not yet distinguishable from converted-photon \sigmapgamma decays.
Although electron reconstruction is more difficult, it is expected that the LHCb experiment
could improve on this measurement and perhaps reach the SM level already with Run 2 data. 
Analogously, the LFV decays \sigmapemu could also be searched for with similar sensitivity. 

Owing to the extreme difficulty of reconstructing neutrons, 
the LHCb experiment will most likely not contribute towards the study of the $\Sigma^-$ hyperon, 
barring exotic channels with baryon number violation.

As far as $\Sigma^0$ particles are concerned, 
these do not have a sizeable decay time, due to their electromagnetic decay into $\Lambda \gamma$, 
therefore they would decay at the production vertex in LHCb. 
For this reason while our simplified model could predict their reconstruction efficiency, 
the sensitivity for $\Sigma^0$ decays would be dominated by primary interaction background, 
which would require a full simulation to be understood. 
We therefore do not provide estimates on these sensitivities. We limit ourselves to suggest that LHCb could attempt a first search for the 
$\Sigma^0 \to \Lambda e^+ e^-$ decay, for which no experimental measurement is currently available, 
despite the fact that several authors proposed this decay to study parity violation in strangeness-conserving weak currents~\cite{Mani:1974wt,Abers:1977dc,DOlivo:1980mri}.
In lieu of an experimental measurement the PDG reports a theoretical calculation driven by internal 
photon conversions for an expected branching fraction of about $5\times10^{-3}$~\cite{Feinberg:1958zz}, 
easily reachable by LHCb if background can be controlled.

\subsection{\boldmath Rare decays of $\Lambda$ hyperons}

\begin{figure}[t!]
  \includegraphics[width=0.49\textwidth]{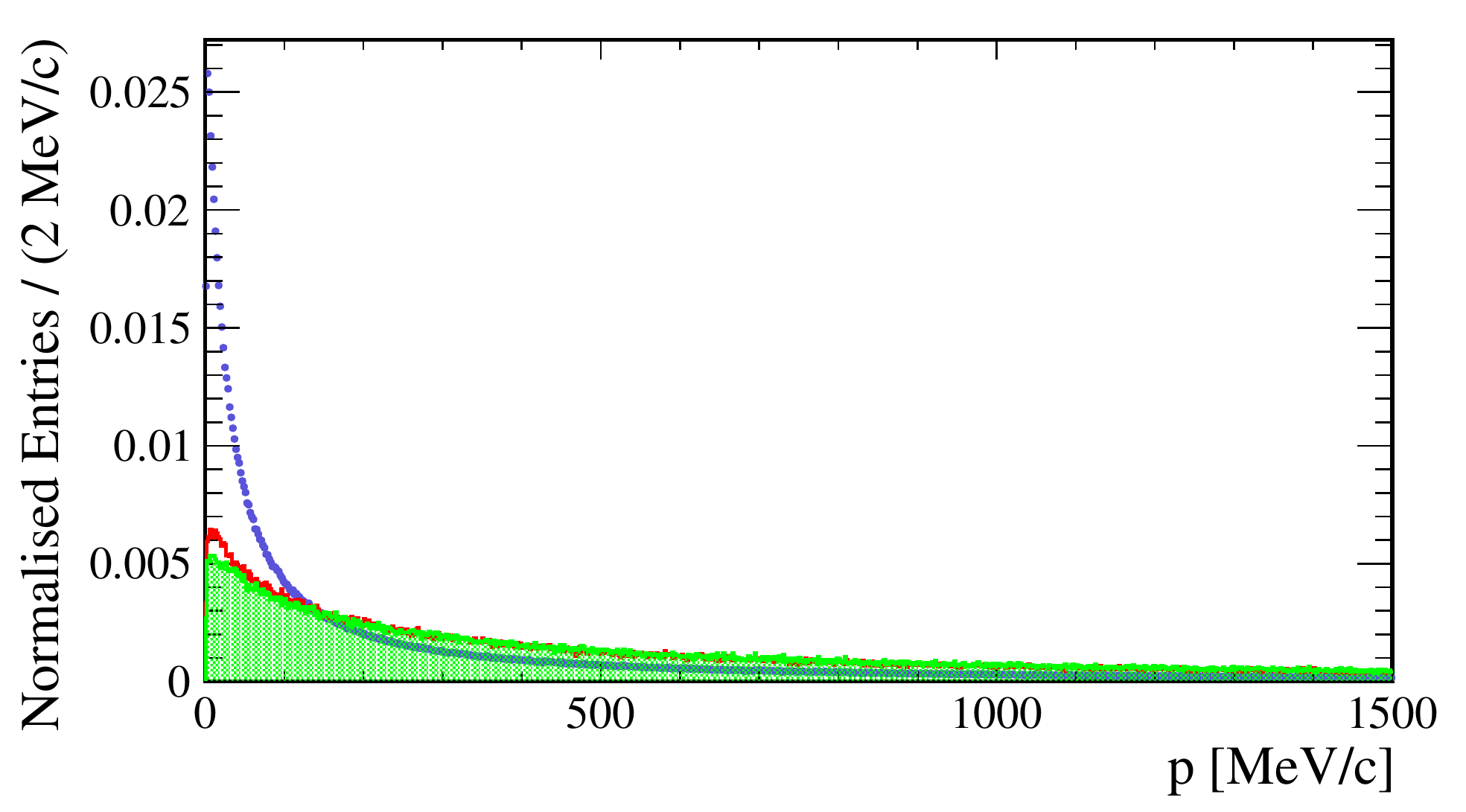}
  \includegraphics[width=0.49\textwidth]{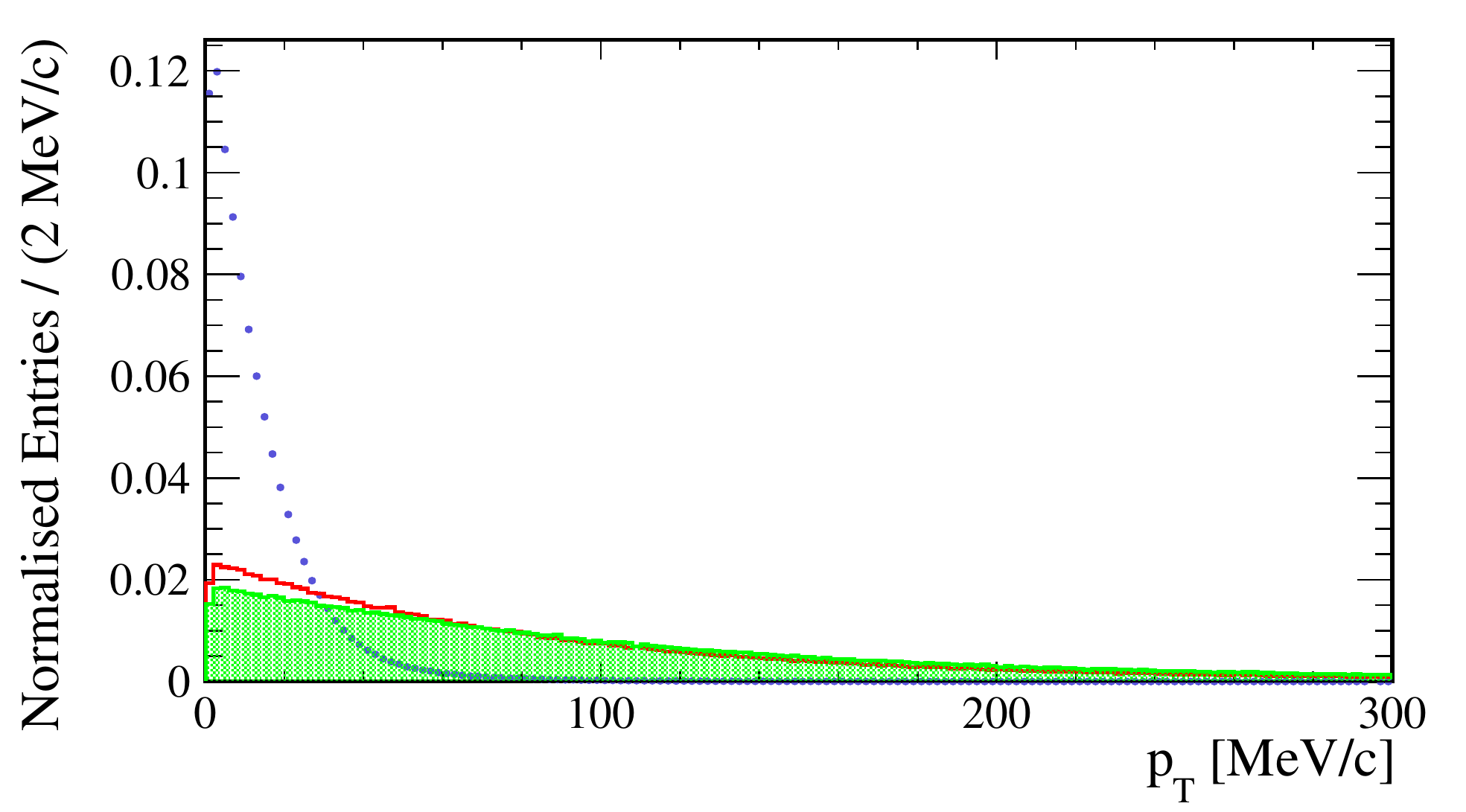}
    \caption{Momentum (left) and transverse momentum (right) for electrons generated in various strangeness decays, where the dotted blue represents \Lppiee, solid red \Kspipiee and filled green \Ksmmee.
    }\label{fig:el_pt} 
\end{figure}

The most compelling contribution LHCb could offer in the realm of $\Lambda$ hyperon is the improvement on the branching fraction 
of the radiative $\Lambda \to p \pi^- \gamma$ decay, 
whose measured value $\mathcal{B}(\Lambda \to p \pi^- \gamma) = (8.4\pm 1.4)\times10^{-4}$,
is known only for pion centre-of-mass momenta less than $95~\mevc$~\cite{Baggett:1973qb}.
In addition, first studies of \Lppiee, which proceeds via flavour-changing neutral currents could be possible, 
reaching branching fractions of $10^{-6} - 10^{-7}$.
A major challenge for \Lppiee is the extremely low transverse electron momentum as illustrated in figure~\ref{fig:el_pt}, translating into a meagre
reconstruction efficiency in accordance with table~\ref{tab:acc_el}.
The corresponding channel with muons in this case would be phase-space forbidden.

LHCb can also advance the study of baryon-number-violating decays, which 
can be produced by virtual particles with masses at the Grand Unified Theory (GUT) scale. For weakly decaying particles, this would imply branching
fractions suppressed proportionally to ${(m_W/\Lambda_{\rm GUT})^4}$, in principle placing observation out of reach for LHCb and any other experiment.
These decays are also indirectly constrained by severe limits from nucleon decays. 
The CLAS collaboration has recently reported searches for several baryon-number-violating $\Lambda$ decays~\cite{McCracken:2015coa}. 
Most of these are in the form $\Lambda \to h \ell$, where $h$ is a $K^+$ or  $\pi^+$ meson and $\ell=e, \mu$ leptons. 
CLAS then provided the first direct experimental limits on such branching fractions to be in the range $[10^{-7}, 10^{-6}]$. 
LHCb can certainly improve on most of these limits, reaching sensitivities  around the $10^{-9}$ level already with Run 2 data. 

\subsection{Rare decays of hyperons with multiple strangeness}

In addition to hadrons with one strange quark or anti-quark ($|S|=1$), LHCb will also produce a large number of baryons with more strange quarks, 
namely the $\Xi$ and $\Omega$ hyperons. As can be seen from figure~\ref{fig:minbias}, 
the production of $\Xi$ is in the region of charmed mesons, 
while $\Omega$ production is further suppressed, due to the additional strange quark, to the level around the beauty meson.
Nevertheless, this provides a large dataset with which to improve existing measurements on these hadrons.

In the context of rare decays, the main interest for $|S|>1$ hyperons is 
for $\Delta S=2$ transitions, which are practically forbidden in the SM,
with branching fractions of order $10^{-17}$. Potential NP transitions mediated  
by parity-odd low-energy operators may enhance the observed rates
while respecting constraints from $K^0 -\bar K^0$ mixing~\cite{He:1997bs}. 
In this respect, the LHCb experiment has the capabilities to improve the branching fraction of
$\Xi^0 \to p \pi^-$, which has an upper limit of $8.2\times 10^{-6}$ at 90\% CL obtained at the HyperCP experiment~\cite{White:2005br}.
This decay has an experimental signature completely reminiscent of the corresponding $\Lambda$ decay, 
which is selected even without particle identification at LHCb~\cite{Aaij:2018vrk}, making it the ideal calibration sample for $\Xi^0 \to p \pi^-$.
Therefore, there is no doubt that the background to this channel could be rejected with high signal retention. 
Branching fractions of order $10^{-9} - 10^{-10}$ could be reached with LHCb Upgrade data. 

In similar vein, the $\Omega \to \Lambda \pi^-$ decay has an upper limit on the branching fraction 
of $2.9 \times 10^{-6}$ at 90\% CL also placed by the HyperCP experiment~\cite{White:2005br}.
The sensitivity to this channel is again expected to be improved over the current limit
given its clean topology, down to branching fractions of order $10^{-8} - 10^{-9}$. 
Incidentally, the channel $\Xi^- \to p \pi^- \pi^-$, which has an upper limit of only 
$3.7 \times 10^{-4}$ at 90\% CL~\cite{Yeh:1974wv}, will also be easily improved by LHCb, 
similarly to $\Xi^0 \to p \pi^-$, reaching sensitivities of order $10^{-9}$.

\section{Other measurements with strange-hadron decays}
\label{sec:others}

\subsection{\boldmath Measurement of the $K^{+}$ meson mass}
\label{sec:mass}

Due to its superb tracking performance, the LHCb detector is particularly suited for a precision measurement of the charged kaon mass. The current experimental average of the $K^+$ meson mass is $m_{K^+}=493.677 \pm 0.013$ \mevcc\cite{PDG}. The uncertainty is dominated by the disagreement between the two most precise measurements, both performed using kaonic atom transitions \cite{Denisov:1991pu,Gall:1988ei}.
Despite the relatively low acceptance in LHCb, the large production cross section for strange mesons in $pp$ collision allows for a large number of $K^+\to\pi^+\pi^-\pi^+$ candidates to be fully reconstructed with an excellent signal-to-background ratio \cite{Contu:1693666}. The number of fully reconstructed decays occurring within the VELO acceptance is estimated to be of $\mathcal{O}(10^7)/\mathrm{fb^{-1}}$ for $pp$ collisions at $\sqrt{s}=13$ TeV with a relatively good mass resolution of $\lesssim 4$ \mevcc \cite{LHCb-PAPER-2017-049}.
Therefore, the statistical error on the mass is expected to be below $10^{-3}$ \mevcc with the entire LHCb dataset. The main systematic uncertainty, which is expected to limit the final precision, will most likely come from the knowledge of the momentum scale resolution, which is proportional to the Q-value of the decay, $m_{K^+} - 3m_{\pi^{\pm}} \approx 75 \mevcc$. For $K^+\to\pi^+\pi^-\pi^+$, this systematic should be below $0.02$~\mevcc~\cite{Needham:1115072}, making this measurement competitive with the world average.

\subsection{Semileptonic decays}
\label{sec:semileptonics}

The latest results from semileptonic $b\to c$ transitions suggest the possibility of BSM contributions in
 charged-current quark decays breaking Lepton Flavour Universality (LFU)~\cite{Ciezarek:2017yzh}. Hence, it is natural
 to investigate if similar patterns can be found in $s\to u$ transitions.

 \subsubsection{\boldmath Semileptonic \KS decays}
 \label{sec:semiks}
 A search for the $\KS\to \pi^{\pm}\mu^{\mp} \nu$ process, which is as yet unobserved experimentally, could be performed at LHCb. This would be useful as a measurement of LFU when comparing to the well-known $\KS\to \pi^{\pm}e^{\mp} \nu$ decay~\cite{PDG}. Depending on the precision achieved, the measurement of this branching fraction could also be useful in constraining the CKM matrix element $|V_{us}|$~\cite{Moulson:2017ive}. However, LHCb would need  excellent control over the systematics to reach the $<$~1\% level of precision that would be required for such a measurement to be competitive. The most challenging background for this search is expected to arise from the corresponding \KL decay to the same final state. The much larger branching fraction of the \KL decay, $\sim27\%$~\cite{PDG}, compensates the reduction in efficiency due to the longer \KL lifetime, leading to significant yields still deposited within the LHCb acceptance: 
 considering the expected $\KS\to \pi^{\pm}\mu^{\mp} \nu$ branching fraction, $(4.69 \pm 0.05) \times 10^{-4}$~\cite{PDG},
the ratio of \KL to \KS events in this final state in the LHCb acceptance is expected to be about 1.5 (4.5) when using long (downstream) tracks,
without further selection. However, given the precise knowledge of the \KL branching fraction,  $(27.04\pm{0.07})\%$~\cite{PDG},
this contribution could be statistically subtracted leaving only a small systematic uncertainty. 
 
 \subsubsection{Semileptonic hyperon decays}
 Semileptonic hyperon decays have been shown to be sensitive to BSM scalar and tensor contributions~\cite{Chang:2014iba}.
The branching fractions of such hyperon decays, which are copiously produced at the LHC, show uncertainties at the $20\%-100\%$ level leaving vast room for progress. 
For example, $\BRof\Lpmunu=(1.57\pm0.35)\times 10^{-4}$, $\BRof\XiLmunu = 3.5^{+3.5}_{-2.2}\times 10^{-4}$ and   $\BRof\XiSmunu < 8 \times 10^{-4}$ at $90\%$ CL.

\begin{figure}[t!]
  \includegraphics[width=0.49\textwidth]{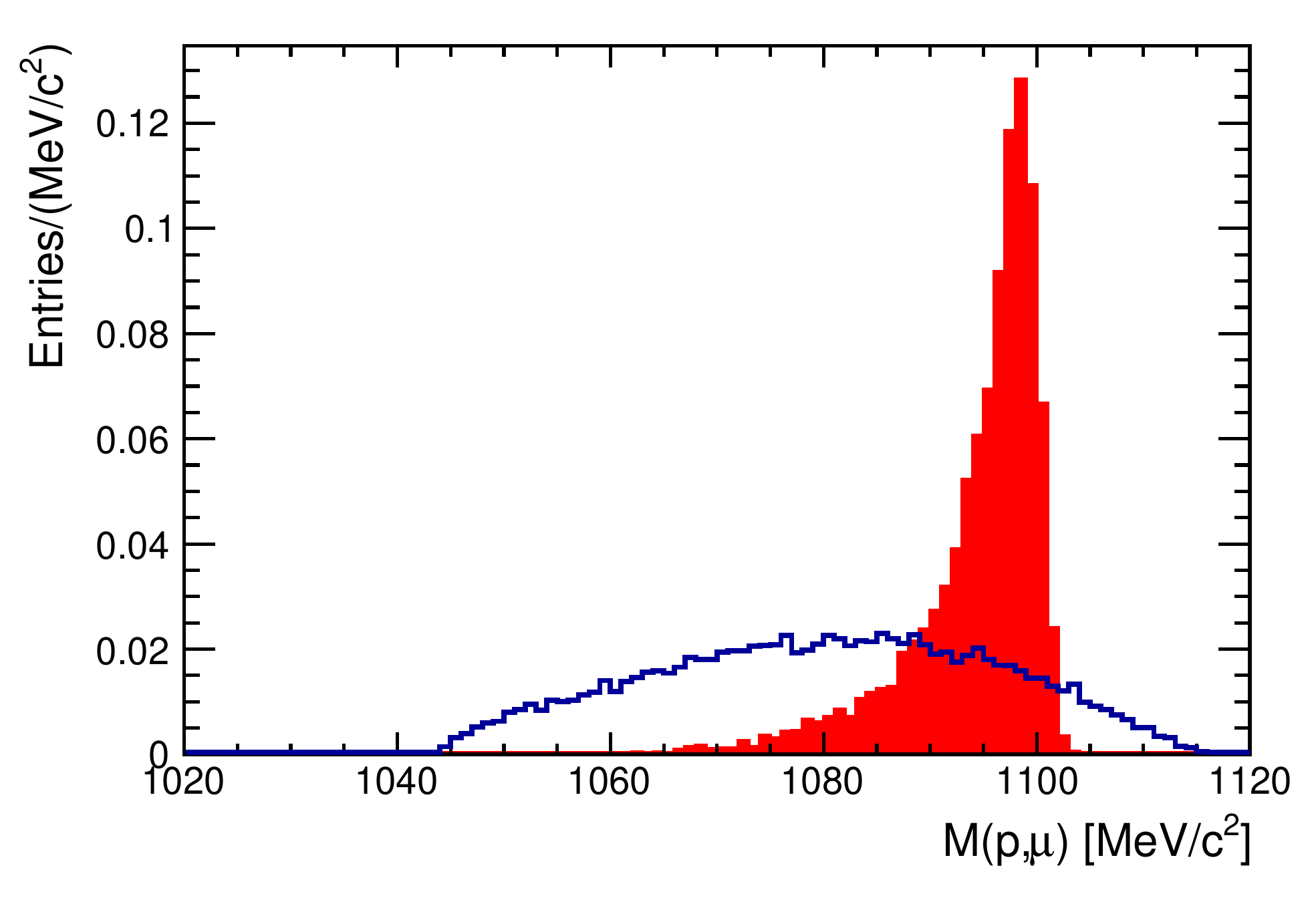}
  \includegraphics[width=0.49\textwidth]{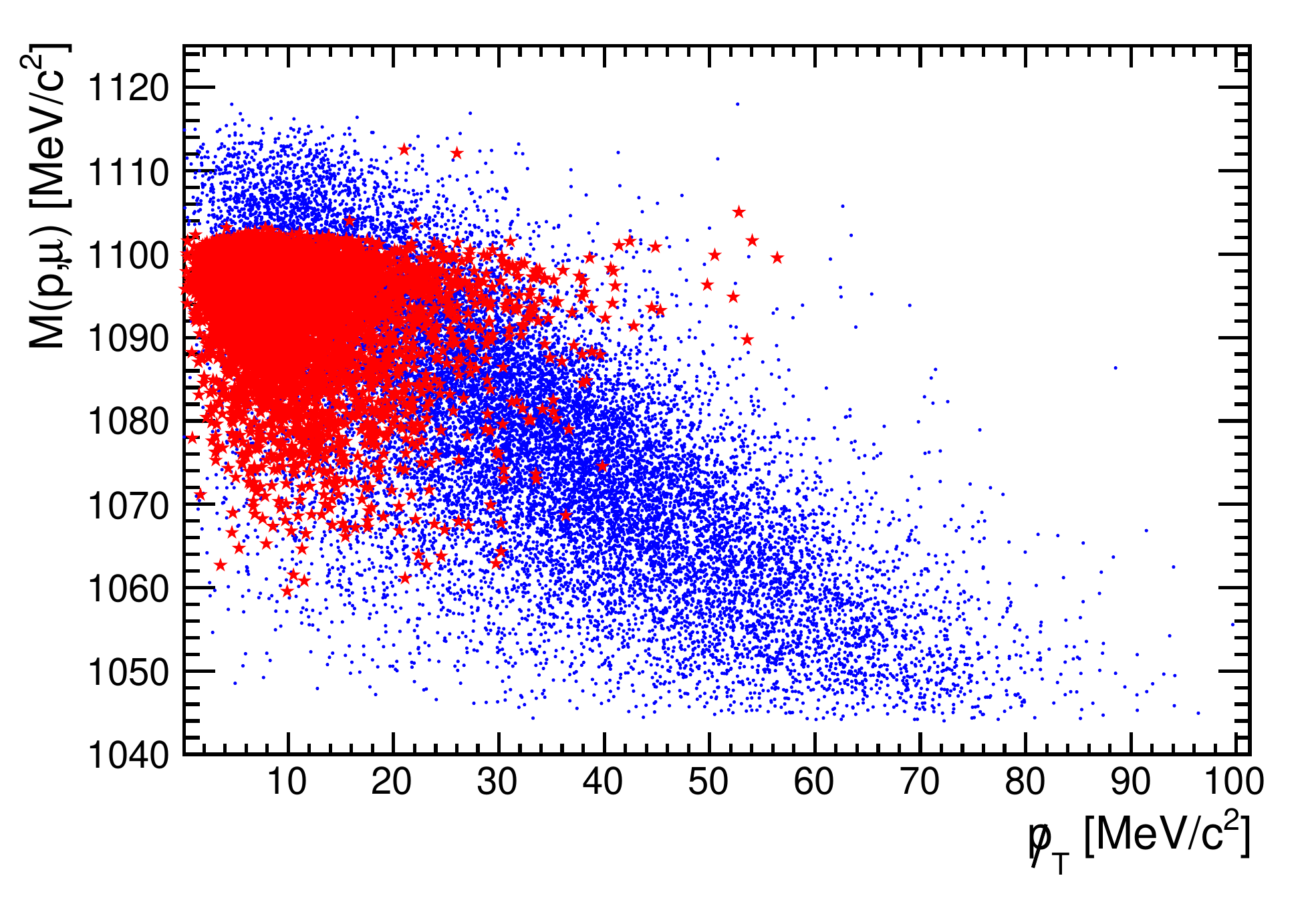}
 
  \caption{\label{fig:Lpmunu} The left plot shows the reconstructed invariant mass for \Lpmunu candidates. Signal events are given by 
    a solid blue line, while the \Lppi background is displayed in filled red.
    The right figure shows a scatter plot of the reconstructed mass {\it vs} missing momentum in the plane transverse to the $\Lambda$ flight direction
  for signal (blue squares) and \Lppi background (red stars). Final state radiation in the $\Lambda$ decay vertex is not included in the simulation.}
\end{figure}

\begin{figure}[t!]
  \includegraphics[width=0.49\textwidth]{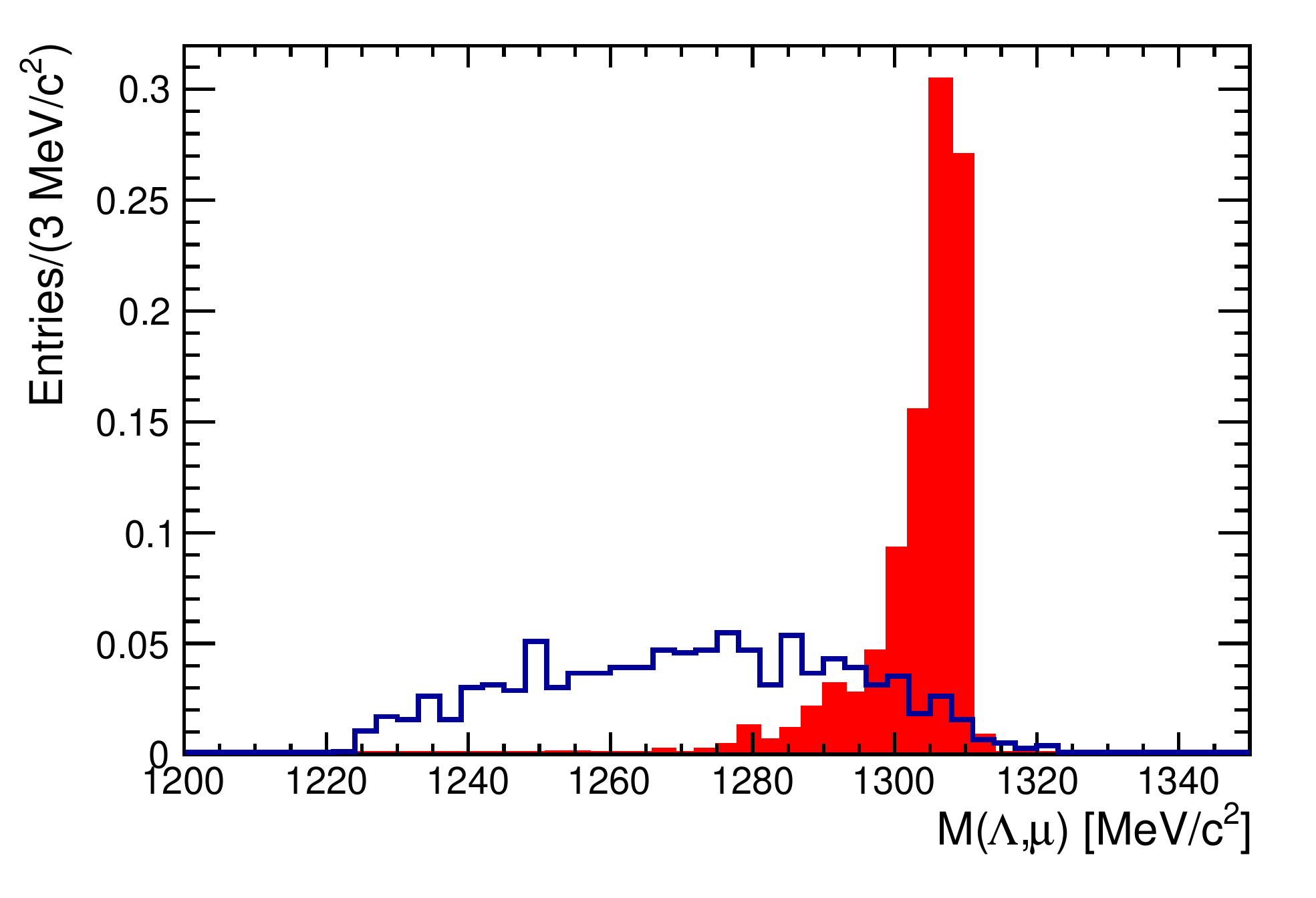}
  \includegraphics[width=0.49\textwidth]{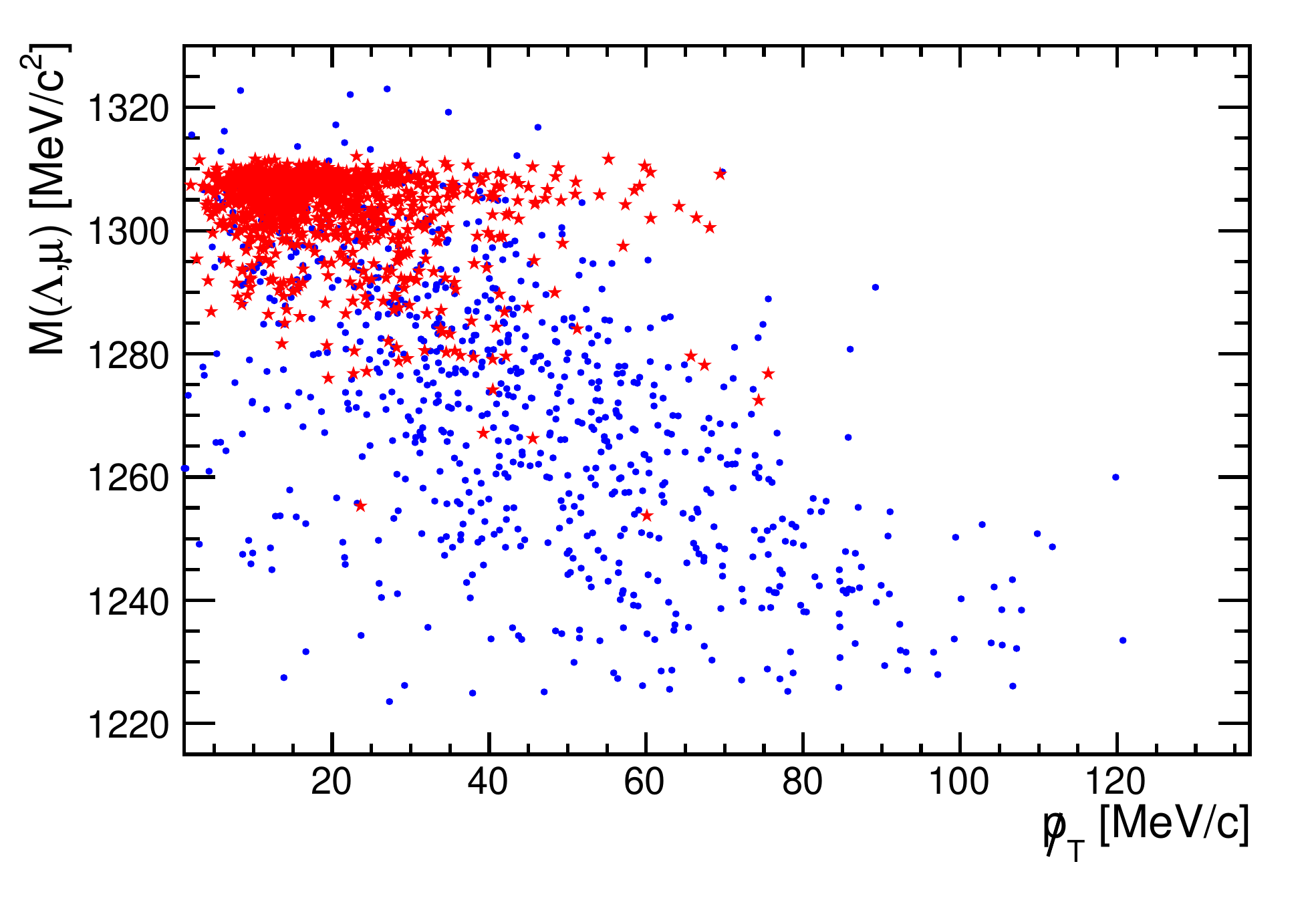}
 
  \caption{\label{fig:XiLmunu} The left plot shows the reconstructed invariant mass for \XiLmunu candidates. Signal events are given by 
    a solid blue line, while the \XiLpi background is displayed in filled red.
    The right figure shows a scatter plot of the reconstructed mass {\it vs} missing momentum in the plane transverse to the $\Xi^-$ flight direction
    for signal (blue squares) and \Lppi background (red stars). Final state radiation in the $\Xi^-$ and $\Lambda$ decay vertices is not included in the simulation.}
\end{figure}

Those decays would be partially reconstructed in LHCb, as was shown in section~\ref{sec:rec}, with improved measurements directly
translating into tighter bounds on LFU, since the electron modes have already been measured very precisely. Kinematic constraints such as those applied in  the
\Kspizmm analysis can be used to reconstruct the strange-baryon peak. Since the expected yields for strange semileptonic decays
are large, the main challenge is not the trigger efficiency, but is instead the discrimination  against peaking backgrounds like \Lppi or \XiLpi.
The mass of the $p,\mu$ candidates from \Lpmunu and misidentified \Lppi is shown in figure~\ref{fig:Lpmunu}, which also plots
the dependency of the mass against the estimated missing momentum transverse to the $\Lambda$ flight direction. Clearly, the signal and peaking background provide contrasting signatures. It has to be noted, however,
that neither final state radiation in the $\Lambda$ decay nor the decay in flight of the pion are included in the simulation,
both of which are effects that can partially dilute the discriminating power of the missing transverse momentum. A similar study is performed
for \XiLmunu, which also demonstrates the separation between signal and the corresponding peaking-background distribution from \XiLpi decays, as depicted in figure~\ref{fig:XiLmunu}.

\section{Competition from other experiments}
\label{sec:competition}

Competition from other experiments on strange-hadron decays will be scarce in the coming years. We briefly review it in the following. 
The NA48 experiment has contributed significantly to the physics of strange-hadron decays, but has already
analysed their full dataset on rare \ks and hyperon decays (e.g. refs~\cite{Batley:2011zza,Batley:2004wg,Batley:2012mi,Batley:2007hp})
and we are not aware of any plan to exploit it further.
The NA62 experiment will give fundamental results on charged kaons, 
however it will not have a neutral beam at its disposal before 2026. 
In particular, NA62 may reach the $10^{-12}$ ballpark in LFV kaon decays~\cite{Petrov:2017wza} with the data collected so far.
The KLOE2 experiment will most probably be able to contribute on semileptonic measurements, in addition to its core \CP-violation program, 
and possibly measure the $K^+$ mass, but will not have enough statistics for rare decays. 
The CLAS experiment could possibly contribute again to searches on rare hyperon decays, 
but will not be competitive with LHCb below the $10^{-7}$ level in branching fraction. 
Similarly it is not expected to contribute on \ks decays. 
Finally, flavour factories such as BESIII and BelleII can possibly contribute to the physics 
of rare strange-hadron decays. The BESIII collaboration has for example published a search 
for $\eta^\prime \to K \pi$ decays~\cite{Ablikim:2016bjc}, reaching a branching fraction limit of order $10^{-4}$.
We are not aware of any published physics result from the Belle collaboration on rare strange hadron decays 
and this topic is not mentioned in BelleII physics book~\cite{Kou:2018nap}.
In addition, there are new proposed facilities such as TauFV~\cite{TauFVTalk}
which may be able to reach ${\cal O}(10^{19})$ kaons in 
the decay volume, with a detector layout comparable to that of LHCb, for which however  
we are not aware of more in-depth sensitivity studies 
on the decay modes discussed in this paper. 
However, we would welcome an increase in the interest for strange physics and
would consider competition from these collaborations to be a very healthy development indeed.

\section{Conclusions}
\label{sec:conclusions}

The decays of strange particles become increasingly important as the energy scale for dynamics beyond the
Standard Model increases. The LHCb experiment has provided the world's best measurements in 
\Ksmm and \sigmapmumu decays with excellent prospects for expanding its research program on strangeness decays.
For the first time, this paper reports estimates of detection efficiencies for several \KS, $K^\pm$ and
hyperon decay channels and evaluates the invariant-mass resolution that could be achieved with the full and downstream tracking systems,
while demonstrating the capacity of LHCb to resolve signal from potential peaking-background distributions.
The results show that several promising new measurements are feasible in various \KS, $K^\pm$ and hyperon decays with diverse
final states.


\section*{Acknowledgements}
{\justifying The LHCb authors would like to thank their collaboration colleagues who participated in helpful discussions on the subject of this paper.
We would like to thank Jorge Portol\'es for helpful discussions in \Kspizmm.
The work of IGFAE members is supported by ERC-StG-639068 ``BSMFLEET'' and XuntaGal. In addition, the work of XCV is also supported by MINECO through the Ram\'on y Cajal program RYC-2016-20073.
JMC acknowledges support from the Spanish MINECO through the Ram\'on y Cajal program RYC-2016-20672. The work of AS is funded by STFC grant ST/J00412X/1.
The work of VVG is partially supported by ERC-CoG-724777 ``RECEPT''.
The work of FD is partially supported by the Science and Technology Facilities Council 
grant ST/N000331/1. AP acknowledges support by SNF under contract 168169. \par}

\addcontentsline{toc}{section}{References}
\setboolean{inbibliography}{true}
\bibliographystyle{fd}
\bibliography{main,LHCb-PAPER}

\end{document}